\documentclass[journal,twocolumn, web]{IEEEtran}
\usepackage{cite}
\usepackage{amsmath,amssymb,amsfonts}
\usepackage{algorithmic}
\usepackage{graphicx}
\usepackage{algorithm,algorithmic}
\usepackage{hyperref}
\hypersetup{hidelinks}
\usepackage{siunitx}
\usepackage{tipa}
\usepackage{lscape}
\usepackage{textcomp}
\usepackage{booktabs}
\usepackage{multirow}
\usepackage{float}
\usepackage{rotating}
\usepackage{tabularray}
\usepackage{array}
\usepackage{url}
\usepackage{verbatim} 
\usepackage[ddmmyyyy]{datetime}

\def\BibTeX{{\rm B\kern-.05em{\sc i\kern-.025em b}\kern-.08em
   T\kern-.1667em\lower.7ex\hbox{E}\kern-.125emX}}
    
\begin{document}

\title{Interpretable Features for the Assessment of Neurodegenerative Diseases through Handwriting Analysis}

\author{Thomas Thebaud, \IEEEmembership{Member, IEEE}, 
Anna Favaro, \IEEEmembership{Member, IEEE},
Casey Chen,
Gabrielle Chavez, 
Laureano Moro-Velazquez, \IEEEmembership{Member, IEEE},
Emile Moukhebeir,
Ankur Butala, 
and Najim Dehak,\IEEEmembership{Member, IEEE}
\thanks{Submitted for review: \today.}
\thanks{Thomas Thebaud, Anna Favaro, Casey Chen, Gabrielle Chavez, Laureano Moro-Velazquez and Najim Dehak are with the Department of Electrical and Computer Engineering, The Johns Hopkins University, Baltimore MD, USA (e-mail: *@jhu.edu, *$\in$\{tthebau1, afavaro1, cchen247, gchavez5, laureano, ndehak3\}). }
\thanks{Moukhebeir Emile is with the Department of Neurology, The Johns Hopkins University, Baltimore MD, USA (e-mail: emoukheiber@jhmi.edu). }
\thanks{Butala Ankur is with the Department of Neurology and the Department of Psychiatry and Behavioral Sciences, The Johns Hopkins University, Baltimore MD, USA (e-mail: Ankur.Butala@jhmi.edu). }
}

\maketitle

\begin{abstract}
Motor dysfunction is a common sign of neurodegenerative diseases (NDs) such as Parkinson's disease (PD) and Alzheimer’s disease (AD), but may be difficult to detect, especially in the early stages.
In this work, we examine the behavior of a wide array of interpretable features extracted from the handwriting signals of 113 subjects performing multiple tasks on a digital tablet, as part of the Neurological Signals dataset. 
The aim is to measure their effectiveness in characterizing NDs, including AD and PD. To this end, task-agnostic and task-specific features are extracted from 14 distinct tasks. Subsequently, through statistical analysis and a series of classification experiments, we investigate which features provide greater discriminative power between NDs and healthy controls and amongst different NDs. Preliminary results indicate that the tasks at hand can all be effectively leveraged to distinguish between the considered set of NDs, specifically by measuring the stability, the speed of writing, the time spent not writing, and the pressure variations between groups from our handcrafted interpretable features, which shows a statistically significant difference between groups, across multiple tasks.
Using various binary classification algorithms on the computed features, we obtain up to 87\% accuracy for the discrimination between AD and healthy controls (CTL), and up to 69\% for the discrimination between PD and CTL. 
\end{abstract}




\newcommand{\STAB}[1]{\begin{tabular}{@{}c@{}}#1\end{tabular}}

 \vspace{-4mm}
\section{Introduction}
\label{sec:intro}


Neurodegenerative diseases (NDs), including Parkinson’s disease (PD) and Alzheimer’s disease (AD), progressively impair cognitive, motor, and behavioral functions by affecting both overlapping and distinct brain regions. 
To monitor these dysfunctions, many studies have explored non-invasive techniques using wearable or digital sensors. 
Among them, handwriting stands out as a promising modality due to its intricate integration of cognitive, motor, and perceptual processes~\cite{thomas2017handwriting, thebaud2023handwriting}.
Handwriting production recruits a distributed neural network across cortical and subcortical structures, including regions involved in learning, executive function, praxis, vigilance, and motor coordination—spanning the cerebral cortex, basal ganglia, and cerebellum~\cite{werner2006handwriting}. Consequently, abnormalities in handwriting patterns can serve as sensitive digital biomarkers of disease presence and progression~\cite{unlu2006handwriting}, with micrographia being a well-established early symptom of PD~\cite{rosenblum2013handwriting}.


Both paper-based and digital handwriting tasks have been employed to evaluate ND-related impairments by analyzing the spatial and temporal characteristics of handwriting signals~\cite{unlu2006handwriting, rosenblum2013handwriting, thomas2017handwriting}.
Historically, most studies have focused on PD, largely due to the overt motor symptoms associated with the disease, including micrographia~\cite{drotar2016evaluation, drotar2014decision, drotar2014analysis, mucha2018identification}, tremor~\cite{smits2014standardized}, and bradykinesia~\cite{smits2014standardized}—collectively referred to as PD dysgraphia~\cite{thomas2017handwriting}.
Handwriting analysis has also been explored for AD detection~\cite{fernandes2023handwriting}, as the disease’s prevalence continues to rise~\cite{cilia2022diagnosing, garre2017kinematic, pirlo2015early}.

Numerous handwriting datasets have been introduced to study NDs, capturing either static images or dynamic time-series signals. However, most datasets are restricted to a single disease type and a single data modality, thereby limiting their experimental scope. 
For the assessment of PD:
\begin{itemize}
\item \textbf{HandPD}~\cite{pereira2015step}: Includes 92 participants (74 PD, 18 CTL), with image-based recordings of spiral and meander tracing. Naive Bayes classification achieved up to 78.9\% accuracy.
\item \textbf{NewHandPD}~\cite{pereira2016deep}: Expanded the previous dataset by adding pressure, acceleration, and tilt data from 14 PD and 21 CTL participants. Using CNNs, the study reached 83.77\% accuracy on spirals and 87.14\% on meanders.
\item \textbf{PaHaW}~\cite{drotar2016evaluation}: Collected position and pressure data from spirals and repeated text in 37 PD and 38 CTL subjects. Using interpretable features and SVMs, it achieved 81.3\% accuracy.
\item \textbf{ParkinsonHW}~\cite{zham2017distinguishing}: Recorded dynamic handwriting from 27 PD and 28 CTL subjects via a digital tablet. Cascaded non-neural classifiers achieved 84.67\% accuracy on spirals and 90.91\% on stability tasks~\cite{sarin2023three}.
\end{itemize}

For the assessment of AD:
\begin{itemize}
\item Early work by~\cite{pirlo2015early} evaluated handwriting signatures from 29 AD and 30 CTL subjects, achieving a 3\% error rate with a decision tree model.
\item The dataset introduced by~\cite{garre2017kinematic} included 23 AD, 12 MCI, and 17 CTL participants performing drawing and writing tasks on a tablet. Sentence copying reached up to 80.4\% classification accuracy across three groups.
\item \textbf{The DARWIN dataset}~\cite{cilia2022diagnosing}, now a standard benchmark, includes 89 AD and 85 CTL participants engaged in copying, memory, and graphical tasks. Fusing non-neural classifiers trained on individual tasks yielded a top accuracy of 94.29\%.
\end{itemize}

Handwriting tasks for ND assessment can be categorized by complexity and task structure:
\textit{Simple motor tasks} involve drawing basic shapes such as spirals or straight lines, or stability exercises, such as maintaining a fixed pen over a tablet. These evaluate tremor, trajectory smoothness, speed, and acceleration~\cite{impedovo2018dynamic}.
\textit{Written language tasks} involve producing words or sentences. On tablets, these tasks enable the study of in-air movement and motor planning between words~\cite{impedovo2018dynamic}. In-air time reflects cognitive preparation~\cite{rosenblum2013handwriting}, and since in-air and on-surface dynamics are non-redundant~\cite{miler2019analysis}, both provide complementary information. Increased in-air duration may reflect hesitation or planning deficits.
\textit{Complex visuoconstructive tasks} include copying figures or drawing clocks. The clock drawing test is an established tool for evaluating visuospatial and executive function across several NDs~\cite{allone2018cognitive, de2010new}. Copying tasks allow examination of written response fidelity in varied contexts~\cite{cilia2019using}.


In this work, we introduce a new dataset composed of digital handwriting samples from 103 participants, including individuals diagnosed with AD, Mild Cognitive Impairment (MCI), PD, and Related Disorders (a heterogeneous group we refer to as \textbf{PDM}, comprising atypical Parkinsonism and misdiagnosed cases).
Participants completed 14 tablet-based handwriting tasks specifically designed to probe cognitive and motor deficits across these conditions.
We extract a large set of interpretable, quantitative measurements, referred to as handwriting features, from these tasks. These include both:
\textit{Task-agnostic features}, which capture global signal properties across all tasks, regardless of their nature, and \textit{Task-specific features}, which target features tailored to particular task types.
We then assess which features best distinguish between groups through statistical analyses and binary classification experiments. In particular, we compare ND subgroups (PD, AD, PDM) to healthy controls, and contrast PD and PDM to identify discriminative features within motor phenotypes.

The main contributions of this paper are as follows:
\begin{itemize}
    \item Introduction of a novel, multi-condition handwriting dataset for ND assessment.
    \item Inclusion of multiple disease groups, including an active control group (PDM), expanding beyond the narrower focus of prior datasets.
    \item Extension of handwriting analysis from PD to AD and MCI, which have received limited attention.
    \item Design and evaluation of a broad set of task-agnostic and task-specific features that align with expected behavioral patterns and reveal significant group differences.
    \item Novel exploration of inter-task comparisons to analyze behavioral shifts when task parameters change (e.g., drawing hand).
\end{itemize}

The sections of the manuscripts are organized as follows: Section \ref{sec:material} presents the data collection pipeline and the main pre-processing stages.
Section \ref{sec:methods} introduces the task-agnostic features extracted from all tasks and the task-specific features designed for specific groups of tasks. Then, various methods used to gauge the significance of our features in the various tasks analyzed are described.
Finally, we introduce the methods used for training classifiers for NDs assessment from the presented interpretable features.
Section \ref{sec:results} shows and discusses our statistical and classification results.
Section \ref{sec:conclusion} contains the main findings, discussions, limits and conclusions of the study.


 \vspace{-4mm}
\section{Materials}
\label{sec:material}
In this section, we outline the data collection and preprocessing methods and the tasks evaluated in this research.

\vspace{-4mm}
\subsection{Data Collection}
\label{subsec:dataset}

The authors of this study collected a dataset from 113 participants from a larger digital biomarker cohort, NeuroLogical Signals, which our group has previously reported \cite{favaro24_odyssey, favaro2024analyzing, favaro2023evaluation, favaro2023interpretable, favaro2023multi, favaro2023multilingual,wangexploring}. These participants were either categorized as healthy (CTL) or diagnosed and treated for a ND by a subspecialist neurologist or geriatric psychiatrist from Johns Hopkins Hospital. Participants diagnosed with NDs were categorized into aforementioned groups based on disease specific clinical diagnostic criteria diagnosis:
\begin{itemize}
    \item AD participants were enrolled from a University AD Center of Excellence, treated by an expert geriatric psychiatrist with an AD diagnosis based upon 2011 criteria~\cite{mckhann2011diagnosis}. However, most of these subjects were enrolled in serum or CSF biomarker studies confirming their diagnosis reflecting later revisions to diagnostic criteria~\cite{bieger2024influence}. 
    \item MCI participants were characterized by established criteria, not yet meeting multidomain impairment or disability consistent with AD~\cite{albert2013diagnosis}. 
    \item PD participants met international recognized diagnostic criteria at a "clinically established"~\cite{postuma2015mds} level of certainty and were followed by a movement disorders neurologist.
    \item PDM participants presented with one or more core symptoms of PD, meeting a "Probable PD" diagnosis but which was later revised based on evolving characteristics or supportive testing. This is a mixed, active comparator group includes patients later diagnosed with Dementia with Lewy Bodies, Fahr’s Disease, Ataxia, Corticobasal Syndrome, Corticobasal Degeneration, Cervical Dystonia, Gerstmann Schenker Schyuler Syndrome, Wilson Disease, MSA-Parkinsonian, and Essential Tremor. In this study, this group is considered an active control group against the PD group.
\end{itemize}
All participants signed informed consent, and the data collection was approved by the Johns Hopkins Medicine Institutional Review Board. 
Participants with PD continued their usual pharmacological treatment and took dopaminergic medication before the recording session.
Handwriting samples were collected using a Wacom One 13 tablet\footnote{\url{https://estore.wacom.com/en-us/wacom-one-13-touch-dth134w0a.html}} with the associated pen. This setup enabled us to record the pressure applied by the pen on the tablet surface and its position both on the tablet and in the air when in close proximity.  Note the recording session also included collection of saccadometric and synchronized speech, although these two modalities were not utilized in this study.
Participants completed 14 written tasks per session, detailed in Section \ref{ssec:tasks}, guided by a research assistant.
We collected recordings for each participant for at least one session, while 21 participants were recorded up to three times, when available and interested, with each session spaced six months apart.
The handwriting data collected is part of a larger, multimodal dataset that is yet to be published.
The distribution of participants, with their age, gender, and total number of tasks recorded, is presented in Table \ref{tab:data}. 
Participants classified as AD and MCI were combined into a single group, named AD* in this study.
Only a subset of the recorded participants were kept in that study, to balance as much as possible the age and gender of the participants.
However, given the low number of female participants, gender balancing for the AD* group was not possible, so we chose to keep the whole group.
\begin{table}[t]
     
    \centering
    \caption{Distribution of the participants recorded in the dataset. }
    \resizebox{\linewidth}{!}{
    \begin{tabular}{c c c c c c c}
\toprule
        	\multirow{2}{*}{Category} & Number of & \multirow{2}{*}{\% Females} & \multirow{2}{*}{Age} & Files &  \multirow{2}{*}{MoCA~\cite{julayanont2017montreal}} & \multirow{2}{*}{UPDRS3~\cite{movement2003unified}}\\
         & Participants & & & recorded & & \\
        	\midrule
            
		CTL	& 42	& 57.1\%	& 69y ($\pm 11$)	& 757 	& 25.4 	&  \\
		PD	& 35	& 40.0\%	& 67y ($\pm 9$)	& 608 	& 25.9 	& 24.3 \\
		PDM	& 15	& 46.7\%	& 53y ($\pm 14$)	& 283 	& 25.8 	& 21.5 \\
		AD	& 21	& 19.0\%	& 70y ($\pm 7$)	& 745 	& 19.7 	&   \\
	\midrule
		Total	& 113	& 43.4\%	& 66y ($\pm 12$)	& 1840 	&   	&   \\
	\bottomrule
    \end{tabular}
    }
    \label{tab:data}
     
\end{table}

\vspace{-4mm}
\subsection{Handwriting Tasks}
\label{ssec:tasks}
\begin{figure}[ht]
    \centering
    \includegraphics[width=\linewidth]{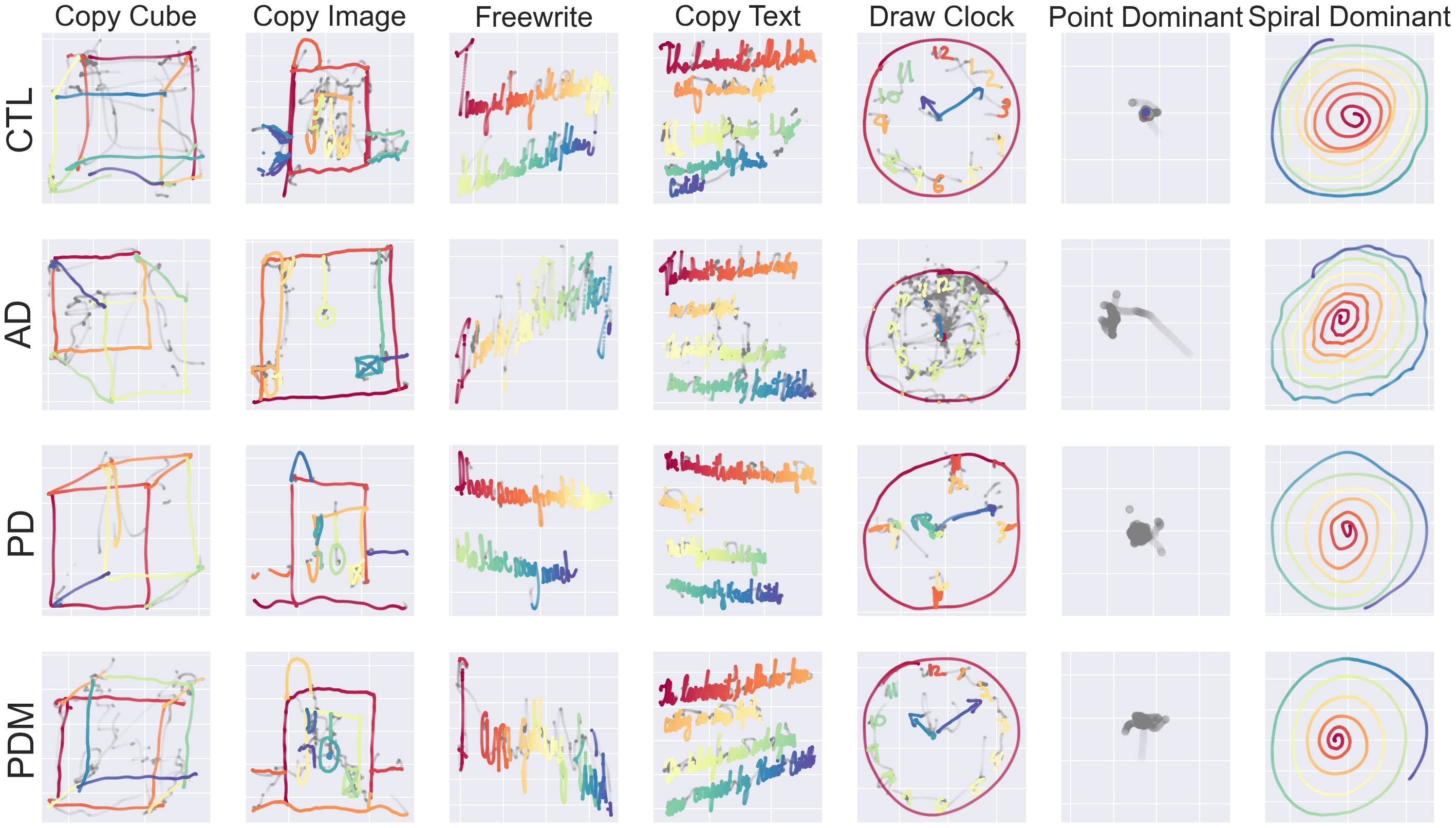}
    \caption{Example of each type of task from all four experimental groups. When on tablet, the color of the points transits from red to purple. When in air, the points are depicted in gray.} 
    \label{fig:examples-tasks}
\end{figure}

In each session, we collected data from 14 handwriting tasks per participant.
Due to myriad reasons, technical or human, the recordings of some tasks can fail, leading to a variable number of participants for each task.
Participants having one or more missing tasks are still included in the study.
Within this section, we provide a comprehensive description of tasks, their origin (what each of those tasks were used for, initially), and outline the experimental hypotheses we formulated for each. 
Tasks such as Point and Spiral tasks involved participants performing stability exercises three times: alternating hands, then with their dominant hand while concurrently speaking. 
\textbf{Writing} and \textbf{Drawing} tasks prompted participants to replicate an item from the screen, from memory, and generate new content.
Participants were asked not to touch the tablet with their arm or hand while writing. 
Figure \ref{fig:examples-tasks} illustrates a sample from each task type, encompassing participants from the CTL, AD, PD, and PDM groups.

\subsubsection{Point Tasks}
The point tasks are designed to assess the static motor control and stability of the participants.
Each participant was asked to maintain the pen vertically above a point shown in the tablet without moving or touching the tablet for \SI{10}{\sec}.
More specifically, in the \textbf{Point Dominant} and \textbf{Point Non-Dominant} tasks, participants are asked to use their dominant and Non-Dominant hand, respectively, and in the \textbf{Point Sustained}, they are asked to use their dominant hand, in addition, to sustain the phonation of the vowel \textit{A} during the exercise.

\subsubsection{Spiral Tasks}
The spiral tasks are also designed to assess the dynamic motor control and coordination of the participants.
Several clinical rating scales~\cite{elble2017digitizing, hernandez2024whiget} require participants with PD to perform regular circular movement on paper or on tablet to assess their regularity or the presence of an eventual tremor~\cite{pereira2018handwritten, saunders2008validity, stanley2010digitized}.
The spiral test has also proven useful for AD assessment using a tablet~\cite{carfora2022extracting}.
The participants were asked to draw a spiral starting from the center of the tablet.
More specifically, in the \textbf{Spiral Dominant} and \textbf{Spiral Non-Dominant} tasks, participants were asked to use their dominant and Non-Dominant hand respectively, and in the dual-task \textbf{Spiral PaTaKa}, they were asked to use their dominant hand, and simultaneously, to perform a spoken diadochokinetic task, i.e., repeat as fast as possible the syllables \textit{pa}, \textit{ta}, and \textit{ka} in this exact order. However, our current focus lies solely on the handwriting data. We hypothesize that the simultaneous performance of tasks will make it more difficult for participants to use strategies to compensate for the motor impairments caused by NDs.

\subsubsection{Writing Tasks}
The writing tasks were designed to assess the cognitive and motor capacities of the participants.
Handwriting analysis has been used mostly for the assessment of PD, as some motor symptoms of the disease can be seen through some characteristic modifications of the handwriting, such as micrographia~\cite{mclennan1972micrographia, letanneux2014micrographia}

Participants first engaged in the \textbf{Copy Text} task, wherein they were tasked with replicating a paragraph of text displayed on the screen. 
We then focus on the impact of multitasking by introducing the \textbf{Copy Read Text} task, where participants copied another text while simultaneously reading it aloud. 
Subsequently, participants' spontaneous writing abilities were captured through the \textbf{Freewrite} task, prompting them to write freely one or two sentences that must not contain any personal information. 
Lastly, we exploring handwriting a different context, the \textbf{Numbers} task, wherein participants were tasked with solving eight simple arithmetic operations summing two natural numbers not exceeding a total of 20, while saying the results of the sums out loud. 
It is worth noting that the Numbers task may be analyzed differently from the writing tasks due to its repetitive nature, necessitating a distinct analytical approach.

\subsubsection{Drawing Tasks}
The Drawing tasks are also designed to assess the cognitive and visuo-constructive capacities of the participants.

\begin{figure}[ht]
    \centering
    \includegraphics[width=0.7\linewidth]{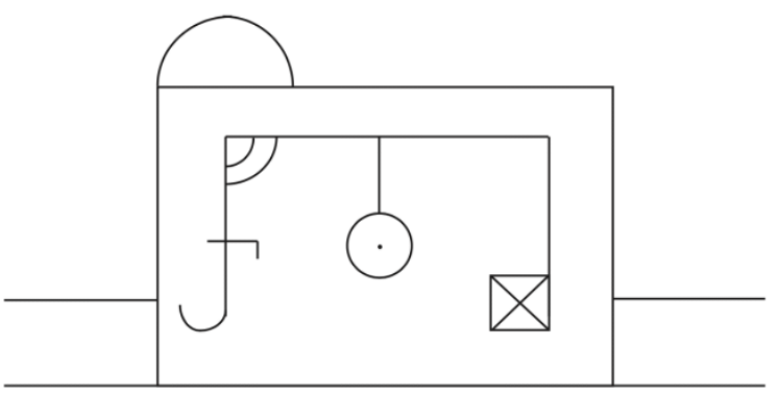}
     
    \caption{Target image presented to the participants for the \textbf{copy image} and \textbf{copy image memory} tasks.}
    \label{fig:copymage_example}
\end{figure}

First, in the \textbf{Copy Image} task, we asked participants to copy an image displayed on the screen.
The image, is a complex figure in the style of those used in the Complex Figure Task~\cite{zhang2021overview}: asymmetric, non-patterned with internal and external details, as shown in the figure \ref{fig:copymage_example}

We further examined the impact of memory on this behavior by introducing the \textbf{Copy Image Memory} task. 
After copying the image in the \textbf{Copy Image} task, participants were then asked to reproduce it from memory approximately five minutes later, thus assessing visual memory and construction abilities. 

Subsequently, participants were asked to perform two tasks: 
First, the \textbf{Draw Clock} task, which looked at their ability to draw all the required elements of a clock at an instructed time~\cite{freedman1994clock}. The Clock Draw Task is another established visuoconstructive screen studied in PD~\cite{de2010new} and AD~\cite{julayanont2017montreal} which has been utilized to distinguish ND~\cite{cahn2003discrimination} and may be interpreted in a number of ways.

Then, the \textbf{Draw Cube} task, used in the MoCA test~\cite{julayanont2017montreal} and which has been proved to be also useful for PD assessment~\cite{bu2013usefulness}, examines their capacity to represent a three-dimensional object.

\vspace{-4mm}
\subsection{Data Curation}
\label{subsec:cleaning}
Every task starts and concludes with a participant crossing an area marked as start or end on the screen, respectively, using a pen. However, the portion of the recording relevant to the analysis is shorter than the entirety captured. Hence, we manually eliminated the empty segments before and after the tasks commenced. Files lacking task-related responses are removed. 

  \vspace{-4mm}
\section{Methods}
\label{sec:methods}

This section outlines the features calculated for each task and the methods used to assess their significance among different experimental groups. 
We developed interpretable features from raw handwriting signals collected during various tasks. 
Our analysis has two aspects: 
first, we examine dynamic handwriting properties common to all tasks;
second, we evaluate performance in specific tasks, such as stability in point tasks and proximity between drawing and target in copying tasks.

\vspace{-4mm}
\subsection{Features computation pipeline}
To compute the features analyzed in this study, we follow the pipeline described in Figure \ref{fig:global_feats_pipeline}. 
Each data file can be described as a time series containing $N\in\mathbb{R}$ points:
\begin{itemize}
    \item The positions $X=(x_i)_{i\in[[1,N]]}$ and $Y=(y_i)_{i\in[[1,N]]}$
    \item The time stamps $T=(t_i)_{i\in[[1,N]]}$
    \item The pressure $P=(p_i)_{i\in[[1,N]]}$
\end{itemize}

We define a clean file as the list $\mathcal{F} = [X_u, Y_u, P_u, T_u]$.\\
We computed various features from the clean files, targeting prioritizing motor, cognitive or behavioral elements. The features can be categorized relative to or irrespective of the task:
\begin{enumerate}
    \item Task-agnostic (see Section \ref{ssec:global_feats}): these features measure primarily dynamic properties of the signal such as speed, noise, and basic ratios.
    \item Task-specific (see Section \ref{ssec:specific_feats}): these features are designed for specific tasks or groups of tasks.
\end{enumerate}

\vspace{-4mm}
\subsection{Task-Agnostic Features}
\label{ssec:global_feats}

\begin{figure}[ht]
    \centering
    \includegraphics[width=\linewidth]{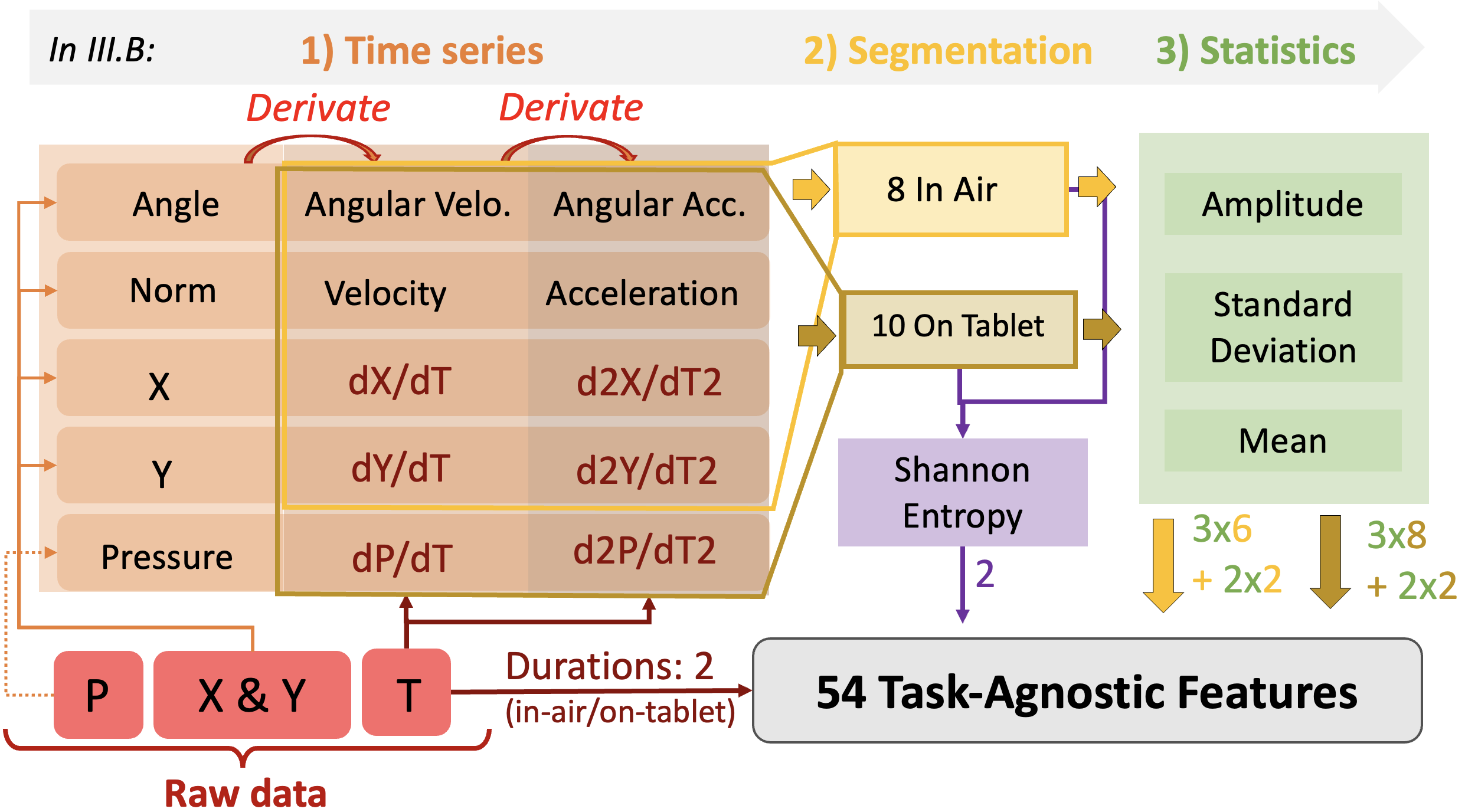}
    \caption{Pipeline used to extract task-agnostic features across all tasks. From the raw data, 5 time series are extracted, differentiated twice, and then the segmentation selects only the in-air parts (yellow path) or the on-tablet parts (brown path), and statistics and entropy are computed from each segmented time-series. The amplitudes for angular velocity and acceleration are not kept, which is why we have 54 and not 58 features.}
    \label{fig:global_feats_pipeline}
\end{figure}

Task-agnostic features quantify the dynamic aspects of the drawings. 
They are computed using several signal-processing techniques.
The extraction process, illustrated in Figure \ref{fig:global_feats_pipeline}, may be parsed:
\begin{enumerate}
    \item computation of time series from the raw data, such as the angle between 2 consecutive points, the distance between two points, and multiple derivations (see Section \ref{sssec:time_series}).
    \item segmentation of each series using only the segments gathered In-Air or On-Tablet. A stroke being defined as the trajectory On-Tablet between two In-Air moments. The segmentation is described in Section \ref{sssec:segmentation}
    \item computation of statistics (e.g., mean, the standard deviation) on the time-series obtained from the derivation of \textit{time independent features}, described in Section \ref{sssec:time_independant_features}.
\end{enumerate}


\subsubsection{Time series}
\label{sssec:time_series}
For every point of a given clean data file, we compute the angle and the $L_2$ norm between 2 consecutive points.
Subsequently, we calculate the first and second derivatives for the following:
\begin{enumerate}
    \item Angle: angular velocity and angular acceleration.
    \item Norm: velocity and acceleration.
    \item Positions: horizontal and vertical velocity and acceleration.
    \item Pressure: pressure variations.
\end{enumerate}
These computations yield a total of 10 columns of data, which are utilized in subsequent analyses.

\subsubsection{Segmentation}
\label{sssec:segmentation}
We implemented segmentation to distinguish between segments occurring on the tablet surface and those in the air. 
Subsequently, we computed time-independent features for the entire file, focusing either solely on points recorded on the tablet or those recorded in the air. 
Given that the \textbf{point task} primarily involves actions in the air, statistical computations are exclusively performed on the in-air segments. Conversely, for spiral tasks, only segments registered on the tablet are utilized. 
However, for writing and drawing tasks, both in-air and on-tablet segments are utilized. 
Notably, features based on pressure are not retained for the in-air segments, as they would be derived from a constant value.

\subsubsection{Time-Independent Features}
\label{sssec:time_independant_features}
From the time series computed previously, we extract a range of statistics to characterize the dynamic properties of the signal, namely: the mean, the standard deviation and the amplitude of the signal.

We chose not to utilize the amplitude measure for angular velocity and acceleration, as it would always yield identical values. 
Furthermore, we calculated the Shannon Entropy based on the points' positions, whether on the tablet or in the air. To achieve this, we employ a Gaussian kernel density estimate (KDE) on the points' $X$ and $Y$ positions. This enabled us to determine the probability $q_i$ for each point $(x_i, y_i)$. Shannon Entropy is then defined as $H_s = -\sum_{i=1}^{N}q_i ln(q_i)$. Following the computation of these features, we obtained a total of \textit{54 task-agnostic features} across the 14 tasks.

\vspace{-4mm}
\subsection{Task-Specific Features}
\label{ssec:specific_feats}
In this section, we detail the various task-specific features we computed. 
We divided the presentations of the features along the four main tasks' categories proposed in Section \ref{ssec:tasks}, with the addition of a section specific to the \textbf{Numbers task}.

\subsubsection{Point Tasks Features}

\begin{figure}[ht]
    \centering
    \includegraphics[width=1\linewidth]{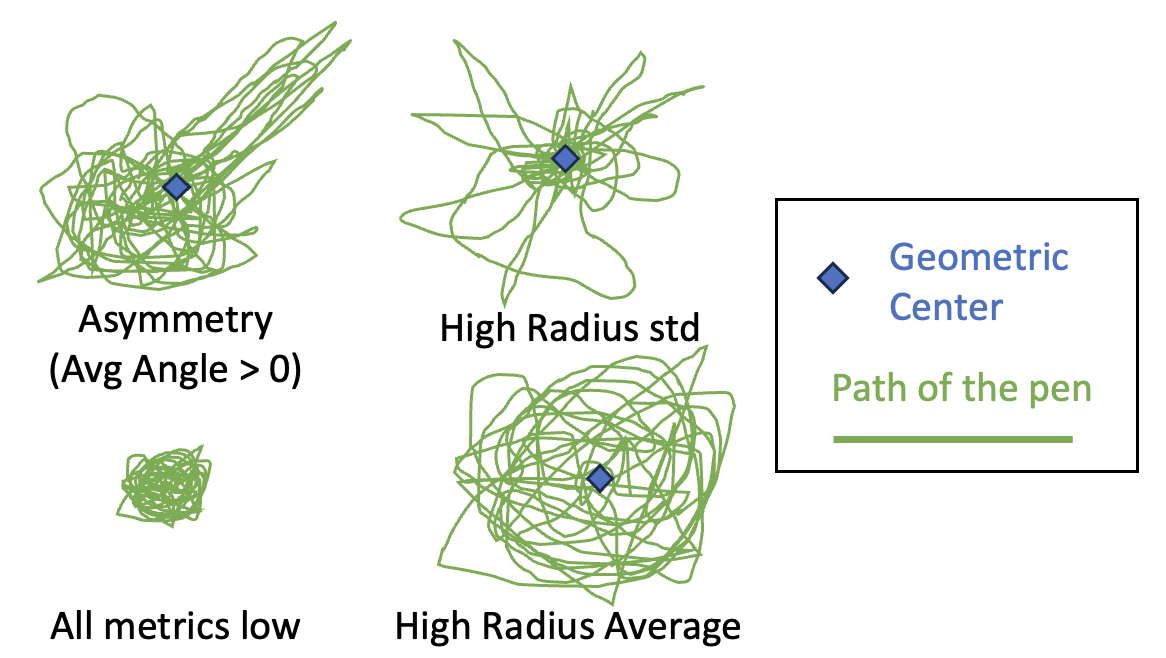}
    \caption{Schematic of the characteristics measured by different point-tasks specific features.}
    \label{fig:point tasks features}
\end{figure}

This section presents the features specific to the \textbf{Point Tasks}. 
Figure \ref{fig:point tasks features} depicts these features.
Point Tasks are mainly used to measure the motor stability of participants and are mostly used for characterization of tremor, primarily relevant to PD or PDM (though tremor may also be observed in AD, due to copathology).
Most dynamic features have already been explored as task-agnostic features, so we focused on the static features of the drawing: how symmetric and regular it is, and what the dispersion of the samples is. 
For this task, we calculated the geometric center of the pen's position over the digital tablet. 
We used this geometric center as a pivot to compare the distribution of the points with respect to the computed center, through 3 features: 
\begin{enumerate}
    \item \textbf{Average Radius} measures the dispersion of pen positions relative to the geometric center, providing a first measurement of stability that assesses the amplitude of variations for essential tremor.
    \item \textbf{Standard deviation of the Radius} measures the variations in the distance from the center, a second measure more focus on dystonic tremors, for irregular vibrations.
    \item \textbf{Average Angle} measures the average of the angle between a point and the center, which can be used as a proxy for the symmetry of the drawing. Tremor and bradykinesia in PD do not always affect all limb regions equally. An average angle $>>$ 0 or $<< 0$ might implicate a strong tremor axis. 
\end{enumerate}

\subsubsection{Spiral Tasks Features}
This section discusses static and dynamic features used to analyze hand-drawn spirals. 

\begin{figure}[ht]
    \centering
    \includegraphics[width=1\linewidth]{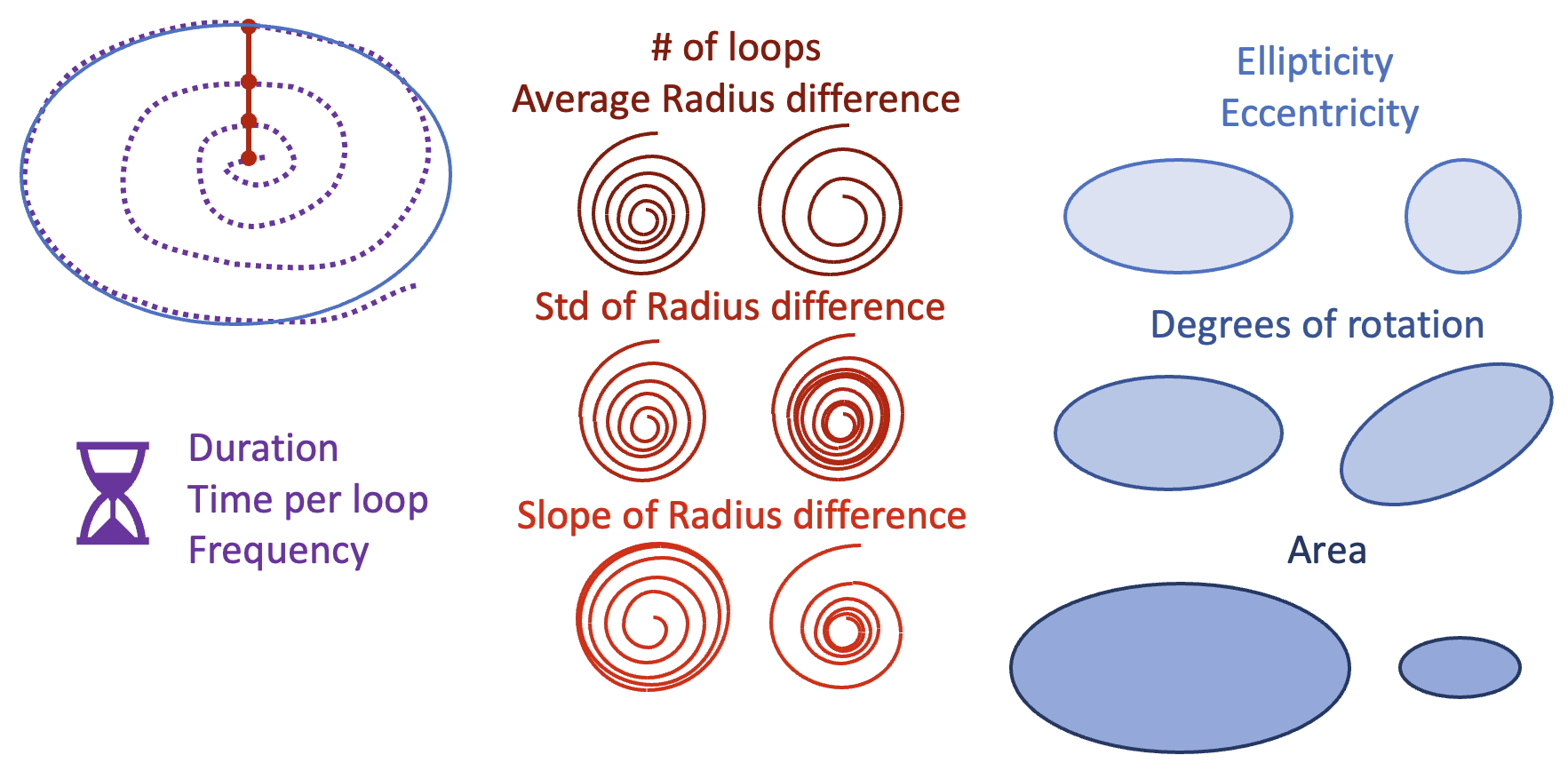}
    \caption{Spiral-specific features computation process}
    \label{fig:schematic-spirals}
     
\end{figure}

Those features are following three directions, illustrated in Figure \ref{fig:schematic-spirals}:
\begin{enumerate}
    \item Static characteristics of the ellipse bounding the spiral.
    \item Static characteristics of the inside loops of the spiral.
    \item Dynamic characteristics of the drawing, based on the time taken. 
    
\end{enumerate}

\paragraph{Ellipse characteristics} 
As shown in Figure \ref{fig:schematic-spirals}, the participants produce rounded spirals for all the spiral tasks. 
Our approach involves initially characterizing the properties of the outer boundaries of the spiral as an ellipse, leveraging these properties as features. 
We used Principal Component Analysis (PCA) to find the directions of the semi-major and semi-minor axes of each spiral, then find the points further away from the geometric center of the spiral in both of those directions to obtain a measure of the semi-minor axis $a$ and the semi-major axis $b$.
Using those values, we define a set of features:
\begin{itemize}
    \item \textbf{Area of the ellipse}: $Area = \pi*a*b$.
    \item \textbf{Eccentricity}: $Ec = \sqrt{1 + \frac{b^2}{a^2}}$.
    The eccentricity is a scalar $\in [0, 1[ $ that measures the compression of an ellipse when defined as a conic section. 
    An eccentricity of 0 would be the one of a circle, while an eccentricity of 1 would indicate a parabola.
    \item \textbf{Ellipticity}: $El = \frac{b}{a}$.
    \item \textbf{Degrees of rotation}: The angle between the semi-major axis and the abscissa.
\end{itemize}

\paragraph{Spiral arms}
Once the outside shape has been characterized, we looked into the properties of the arms of the spiral.
Each spiral is split into segments, containing each one full loop. 
For each loop, we computed its radius as the average distances between its points and the geometric center of the spiral, which allowed us to compute another set of features:
\begin{itemize}
    \item \textbf{Number of loops}
    \item \textbf{The Average distance between loops}: the difference between 2 loops radii is defined as the distance between those 2 loops. We take the \textit{mean} of this measure for all pairs of consecutive loops.
    \item \textbf{The standard deviation of the loops radius differences}: the standard deviation of the radius difference between consecutive loops. This feature quantifies the regularity of the spiral.
    \item \textbf{The slope of the loops radius differences}: the \textit{slope} of the radius difference between consecutive loops.
\end{itemize}

\paragraph{Dynamic analysis}
Dynamic features are then computed based on the total time taken to complete the task:
\begin{itemize}
    \item \textbf{Duration}: total time taken to draw the spiral.
    \item \textbf{Time per loop}: the average time taken to draw one loop.
    \item \textbf{Frequency}: the average number of loops computed per second.
\end{itemize}

\subsubsection{Numbers Task Features}
This section focuses on the features unique to the Numbers task. 
Participants engaged in solving addition problems by moving down the tablet, allowing for individual analysis of each response. 
This methodology enabled us to track the evolution of handwriting features as participants navigate through the task. 
For a comprehensive breakdown of these features, please refer to Figure \ref{fig:numbers tasks features} below.
\begin{figure}[ht]
    \centering
    \includegraphics[width=1\linewidth]{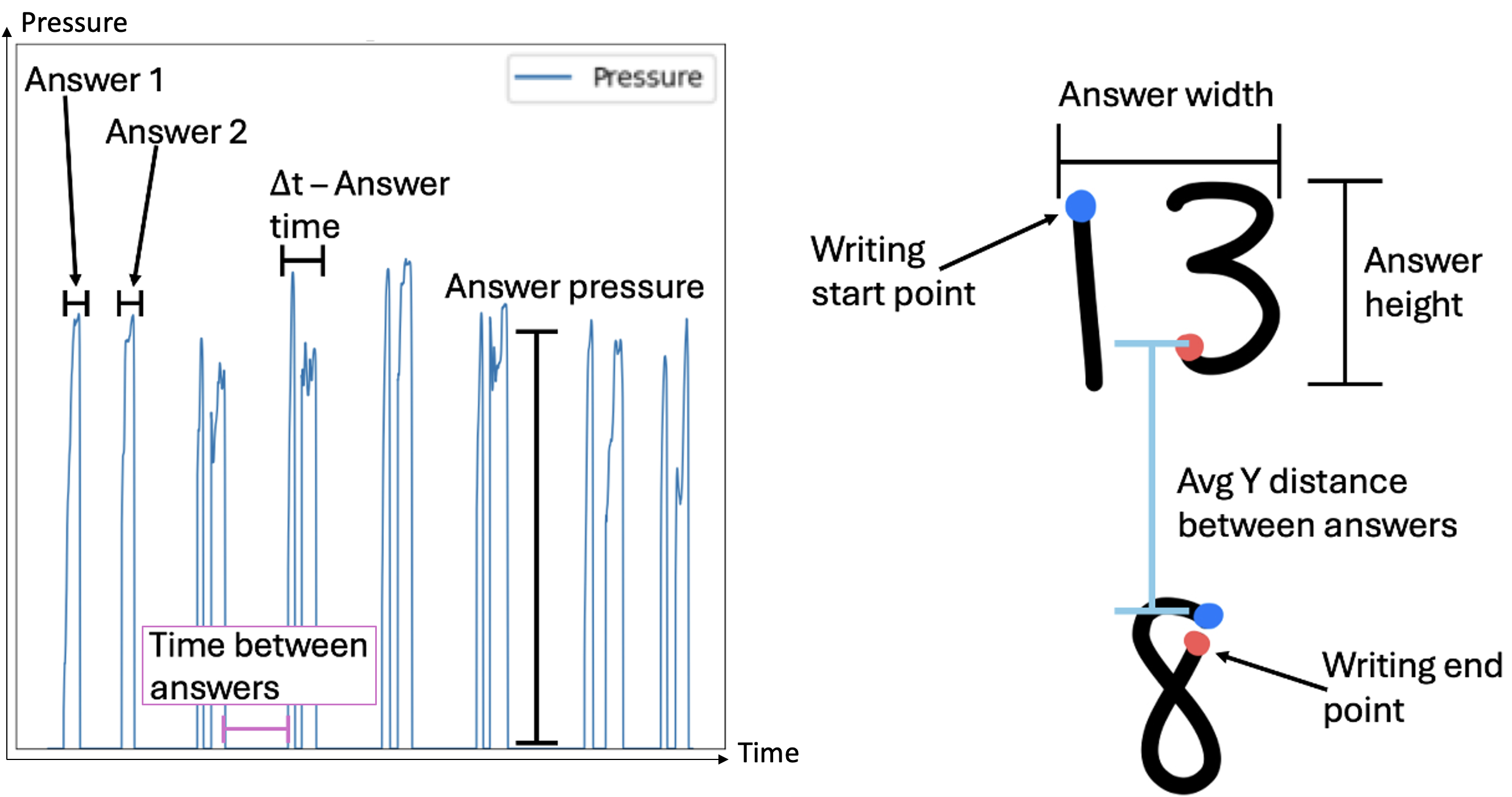}
    \caption{Schematic of the features used for each answer of the Numbers task. The left graph shows the variations in pressures across the whole task, and the right schematic shows the features measured between 2 consecutive answers.}
    \label{fig:numbers tasks features}
\end{figure}

\paragraph{Average Width of Answer} The average width of each answer quantifies the size of the handwriting. It is computed by subtracting the minimum $X$ coordinate from the maximum $X$ coordinate for each answer and then averaging these values.
People with PD have been found to have less control of their movement amplitudes and variance, resulting in compensatory longer strokes and larger handwriting under medication~\cite{van1999parkinsons}, as they are during the studied tasks.

\paragraph{Change in Width of Answer} The variation of width from one answer to the next.
It is computed through a linear regression between the width of each answer and the question number. 
The feature itself represents the slope of this regression line.
The variation in width is aimed to measure the presence of micrographia~\cite{mclennan1972micrographia}.

\paragraph{Change in Average Answer Pressure} The average answer pressure feature is a measure of the amount of pressure that a person applies while writing the answer. This feature can be leveraged as a time series feature to track its variation as the participant progresses through the task. 
It is computed through a linear regression between the average pressure of each answer and the question number. The feature itself represents the slope of this regression line.
This feature might help pick up dystonia features, involuntary muscle contractions commonly seen in neurodegenerative diseases.

\paragraph{Time per Answer} The total duration feature measures the time a participant spends answering one of the questions. Given the absence of a predefined time limit for the Numbers task, this feature provides insight into the participant's pace in completing the task. 

\subsubsection{Writing Tasks Features}
\begin{figure}[ht]
    \centering
    \includegraphics[width=1\linewidth]{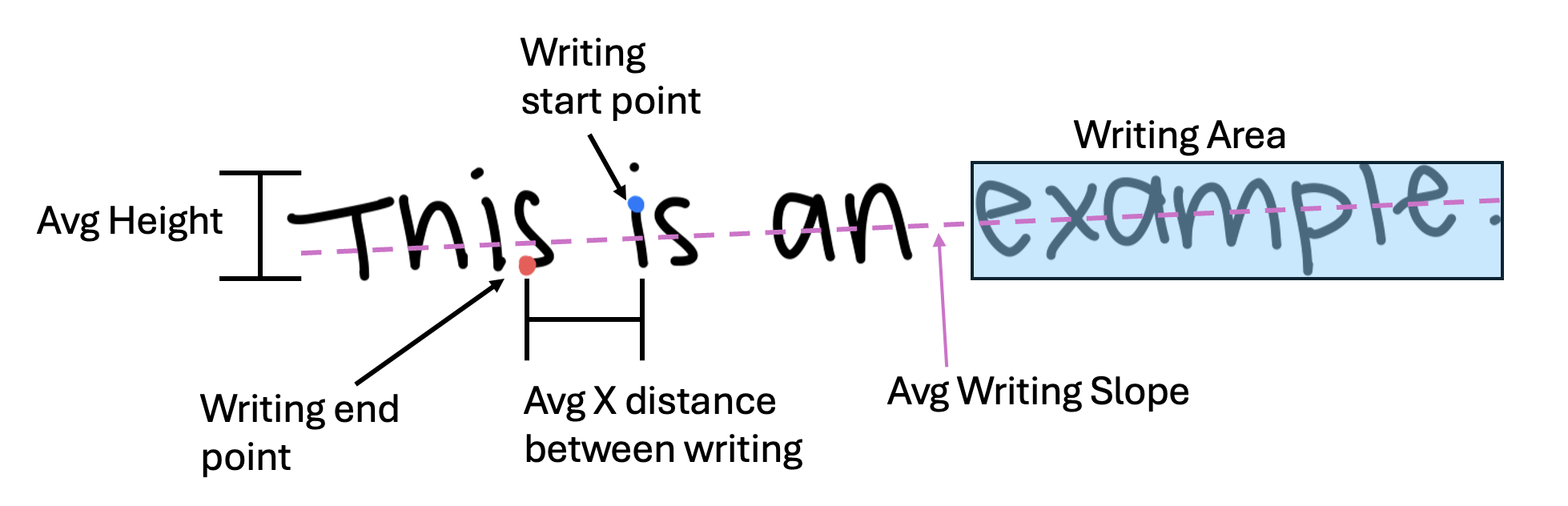}
    \caption{Schematic of the features used in writing tasks, illustrated for the words 'This is an example'.}
    \label{fig:writing tasks features}
\end{figure}

In this section, we present the features extracted from the writing tasks: \textbf{freewrite}, \textbf{copy text}, and \textbf{copy text memory}\cite{chen2024cognitive}. The computation of these features is shown in Figure \ref{fig:writing tasks features}.

\paragraph{Pressure Standard Deviation} The pressure standard deviation feature is a measure of the spread of the amount of pressure a person applies while writing. This measures how steady the hand is when writing by capturing the variations in pressure. Only the nonzero pressure values were considered when calculating this feature, as no pressure can be computed if the pen is not touching the tablet.

\paragraph{Average Time Between Writing} The average time between writing measures how long it takes for the participant to start writing after finishing a word. This shows how fluent the writing is and how stable the motor planning activity is. The feature was calculated by averaging the total time the pen was not touching the tablet.

\paragraph{Average Horizontal Distance Between Words} This feature was calculated by finding the difference between the $X$ value of when the pen is lifted and when it is put back down. Only the $X$ distance was considered because the $Y$ distance between the words is dependent on both the letters they are writing and how they write the letters. 

\paragraph{Total Time} The total time feature gauges the duration required for task completion. It is determined by calculating the time elapsed from when the participant initially places the pen to start writing until they lift the pen after writing the last word. This feature holds greater significance for tasks like Copy Text and Copy Read Text, where participants' required text to write is the same.

\vspace{-4mm}
\subsection{Statistical Analysis}
\label{subsec:stats}
After extracting a total of 76 features (i.e., 54 task-agnostic features and 22 task-specific features), we aim to evaluate their relevance for assessing and characterizing neurodegenerative diseases (NDs). Our objective is twofold: first, to identify which features effectively differentiate the CTL group from each of the disease groups (AD*, PD, and PDM), and second, to assess whether handwriting features can distinguish PD from other forms of Parkinsonism (PDM).

To this end, we perform a Kruskal-Wallis H-test across the four experimental groups (CTL, AD, PD, PDM) for each feature, using the scipy.stats.kruskal function from the Scipy toolkit\footnote{\url{https://docs.scipy.org/doc/scipy/reference/generated/scipy.stats.kruskal.html}}. 
When a participant had multiple sessions, the average of all sessions for each feature is used, to insure the independence of all observations.
For features showing a significant global effect (p$<$0.05), we apply Dunn’s post hoc test to identify specific group differences. To control for false discoveries arising from multiple testing, we apply global False Discovery Rate (FDR) correction across all tasks and features for each comparison, using the statsmodels.stats.multitest.fdrcorrection function\footnote{\url{https://www.statsmodels.org/dev/generated/statsmodels.stats.multitest.fdrcorrection.html}}. To complete our analysis of the separability of those groups of participants, we measured the Area Under the Curve (AUC) for pairs of participant groups, which will be presented alongside FDR-corrected p-values.

Additionally, to investigate how individual features relate to cognitive and motor functioning, we compute the Pearson correlation coefficient between each of the 76 extracted features and both the MoCA~\cite{julayanont2017montreal} (Montreal Cognitive Assessment, for all participants) and UPDRS-III~\cite{movement2003unified} (Unified Parkinson's Disease Rating Scale, motor section, only measured for the PD and PDM participants) scores across participants.
To provide a more detailed analysis of motor function, we also compute correlations between each feature and the individual subdomains of the UPDRS-III score, including:
rigidity, upper extremity bradykinesia, lower extremity bradykinesia, arising from a chair, gait, freezing of gait, postural stability, posture, gait and posture composite, kinetic tremor, postural tremor, resting tremor, global tremor score, and non-tremor components.

Significant results from both the group comparisons and correlation analyses are presented and discussed in the following sections.

\subsubsection{Grouped features}
We hypothesized that certain features extracted for a specific type of task would exhibit consistent behavior across all tasks. To test this hypothesis, we grouped the tasks together. For each group, we conducted the same statistical comparisons outlined in section \ref{subsec:stats}. However, instead of analyzing the recordings from individual tasks, we considered all recordings within each task group. We then measured the significance of all features for each of these task groups as if they were a single task.
\begin{itemize}
    \item \textbf{All spiral tasks}: \textbf{Spiral dominant}, \textbf{Spiral non-dominant}, and \textbf{Spiral PaTaKa}.
    \item \textbf{All point tasks}: \textbf{Point dominant}, Point \textbf{non-dominant}, and \textbf{Point sustained}.
    \item \textbf{All writing tasks}: \textbf{Freewrite}, \textbf{Copy Text}, and \textbf{Copy Read Text}.
    \item \textbf{All drawing tasks}: \textbf{Draw Clock}, \textbf{Draw Cube}, \textbf{Copy Image}, and \textbf{Copy Image Memory}.
\end{itemize}

Table \ref{tab:resume_task features} summarizes the tasks at hand sorted per group and the specific features proposed per group of tasks.

\begin{table}[ht]
 
    \centering
    \caption{summary of the tasks at hand and their specific features.}
    \resizebox{0.95\linewidth}{!}{
    \begin{tabular}{c c r}
        \toprule
        Group & Tasks & Specific Features \\
        
        \midrule
        
               & Point Right        & Radius from centroid \\
        Points & Point Left                 & Angle from Centroid\\
               & Point Sustained          & Std of the Radius \\
        
        \midrule

        & &  Area of the spiral \\
        & Spiral Right &  Eccenctricity \\
        &   & Ellipticity \\
        Spirals & Spiral Left  & Number of loops\\
        &  & Distance between loops \\
        & Spiral PaTaKa  & Time per loop \\
        & &  Frequency \\

        \midrule
        
        &  & Avg width of answer \\
        Numbers & Numbers  & Change in Avg pressure\\
        &  & Time per answer \\
        
        \midrule

        & &  Pressure Ratio \\
        & Copy Text         & Pressure std. \\
        Writing & Copy Text Memory  & Avg time between words \\
        & Freewrite     & Avg distance between words \\
        & &  Total time \\
        
        \midrule

        & Copy Image  \\
        \multirow{2}{*}{Drawing} & Copy Image Memory  & CLIP similarity image-image\\
        & Draw Clock   & CLIP similarity text-image \\
        & Copy Cube     & \\
        
        \bottomrule
    \end{tabular}}
    \label{tab:resume_task features}
     
\end{table}


\vspace{-4mm}
\subsection{Classification Experiments}

In addition to conducting statistical analyses, we conduct binary classification experiments across various experimental groups using features extracted from each task. Namely, it involves a binary classification (e.g., AD* vs CTL) that facilitates the assignment of samples to their corresponding groups, thereby enabling the evaluation of the features' effectiveness in distinguishing between the categories under investigation. As for the statistical analysis, the feature set used in the classification models consists of both task-specific and general features, ensuring a comprehensive representation of handwriting dynamics. The number of features varies across tasks, with Copycube, Drawclock, Copymage Memory, and Copymage containing 54 features, while Copyreadtext, Copytext, and Freewrite include 62 features. The Numbers task has 68 features, whereas Point DOM and Point NONDOM each contain 28 features, and Point Sustained has 27 features. Additionally, Spiral DOM includes 38 features, while Spiral NONDOM and Spiral Pataka each have 37 features. These features capture various aspects of handwriting, including dynamics (e.g., velocity, acceleration, jerk), spatial properties (e.g., trajectory deviations, area), and pressure dynamics (e.g., mean pressure, pressure variability). 

These experiments further evaluate the discriminative power of the features and the utility of tasks in extracting valuable information for characterizing different disorders. We are employing a range of classifiers, including Random Forest (RF), Bagging (BG), and Multi-Layer Perceptron (MLP) \footnote{To implement the different classifiers, we are using the Scikit-learn (Sklearn) library in Python. See \url{https://www.tutorialspoint.com/scikit_learn/scikit_learn_introduction.htm}}. For each classifier, hyperparameter tuning was performed using nested cross-validation (NCV) to ensure optimal generalization performance. Specifically, for the BG classifier, we tuned \textit{max\_samples} (proportion of training samples used per estimator) and \textit{n\_estimators} (number of base estimators). For the MLP classifier, we optimized \textit{hidden\_layer\_sizes} (number of neurons in the hidden layer). Finally, for the RF classifier, we tuned \textit{n\_estimators} (number of trees in the forest) and \textit{max\ features} (number of features considered per split). NCV is a widely used approach for assessing model performance, especially when dealing with limited data, as it helps provide an unbiased estimate of how well the model will generalize to new data. By incorporating two levels of data partitioning, NCV enables both model evaluation and hyperparameter tuning while minimizing overfitting and overly optimistic performance estimates \cite{cawley2010over, tsamardinos2015performance}.
The NCV framework consists of two levels of cross-validation to ensure robust model evaluation and mitigate overfitting. In our implementation, the \textit{outer loop} employs a 10-fold cross-validation, partitioning the dataset into 10 folds, with one fold held out for testing while the remaining nine are used for training. Within each outer training set, an \textit{inner 10-fold cross-validation} is conducted to optimize hyper-parameters. The best hyperparameter configuration is selected based on the \textit{average F1-score} computed across all 10 inner folds, which helps mitigate variance in performance estimates and ensures stability in hyperparameter selection. The F1-score was chosen as the optimization criterion due to class imbalance. To prevent data leakage, cross-validation was strictly performed at the subject level. Speaker-independent folds were created to ensure that all samples from a given individual appeared only in either the training or test set within any iteration. This setup prevents data from the same subject from being in both training and testing, thereby avoiding overly optimistic performance estimates. The entire NCV process involved 100 hyperparameter tuning runs (10 outer folds $\times$ 10 inner folds), with final performance estimates derived by averaging results across the outer folds. This methodology provides a robust assessment of model generalization while maintaining strict speaker independence.
For our chosen model, the Bagging Classifier, the following list of hyper-parameters was explored, and the best combination for each task was kept:
\begin{itemize}
    \item \textit{$max\_samples$}: 0.05, 0.1, 0.2, 0.5
    \item \textit{$n\_estimators$}: 10, 100, 1000
\end{itemize}
The full list of explored hyper-parameters for every explored model is provided in the associated github\footnote{The full code allowing for the reproduction of all experiments will be released upon acceptance on github}.

\begin{figure*}[t]
\center
 
\includegraphics[width=0.9\textwidth]{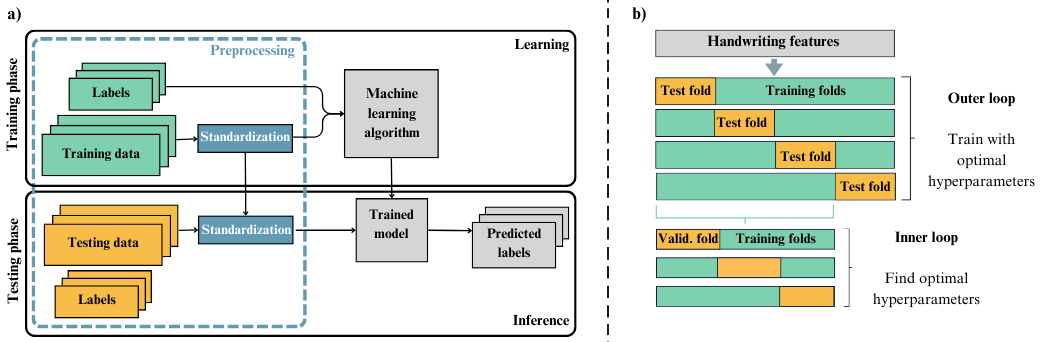}
\label{fig:classif_pipeline}
 
\caption{a) Machine learning pipeline was adopted for training and testing. 
b) Nested cross-validation for algorithm selection. In the graph, the nested cross-validation process is exemplified using 3 folds for training, one for testing in the outer loop, 2 for training, and one for validating in the inner loop. 
However, in our experiments, we used 9 folds for training, one for testing in the outer loop, 8 for training, and one for validating in the inner loop. This strategy was applied separately when training each of the classifiers adopted in the experiments.}
\centering
 
\end{figure*}
 \vspace{-4mm}
\section{Results}
\label{sec:results}
In this section, we first present the most significant features obtained by groups of features for each test realized.
Then, we comment on the results obtained in the classification experiments for each task.

\vspace{-4mm}
\subsection{Task-agnostic Features Results}
Out of 54 task-agnostic features computed, we found 53 were significant ($p\leq0.05$) for, at least one task and one groups comparison.
The complete table is available in the Appendix; however, we prioritize limited sets due to the considerable number of significant features. 
We present the features most effectively distinguishing AD* and PD from control participants and separate PD from PDM participants. 
Subsequently, we will highlight features that exhibited robust behavior across multiple comparisons.

\subsubsection{AD* vs CTL}
\begin{figure}[ht]
    \centering
    \includegraphics[width=\linewidth]{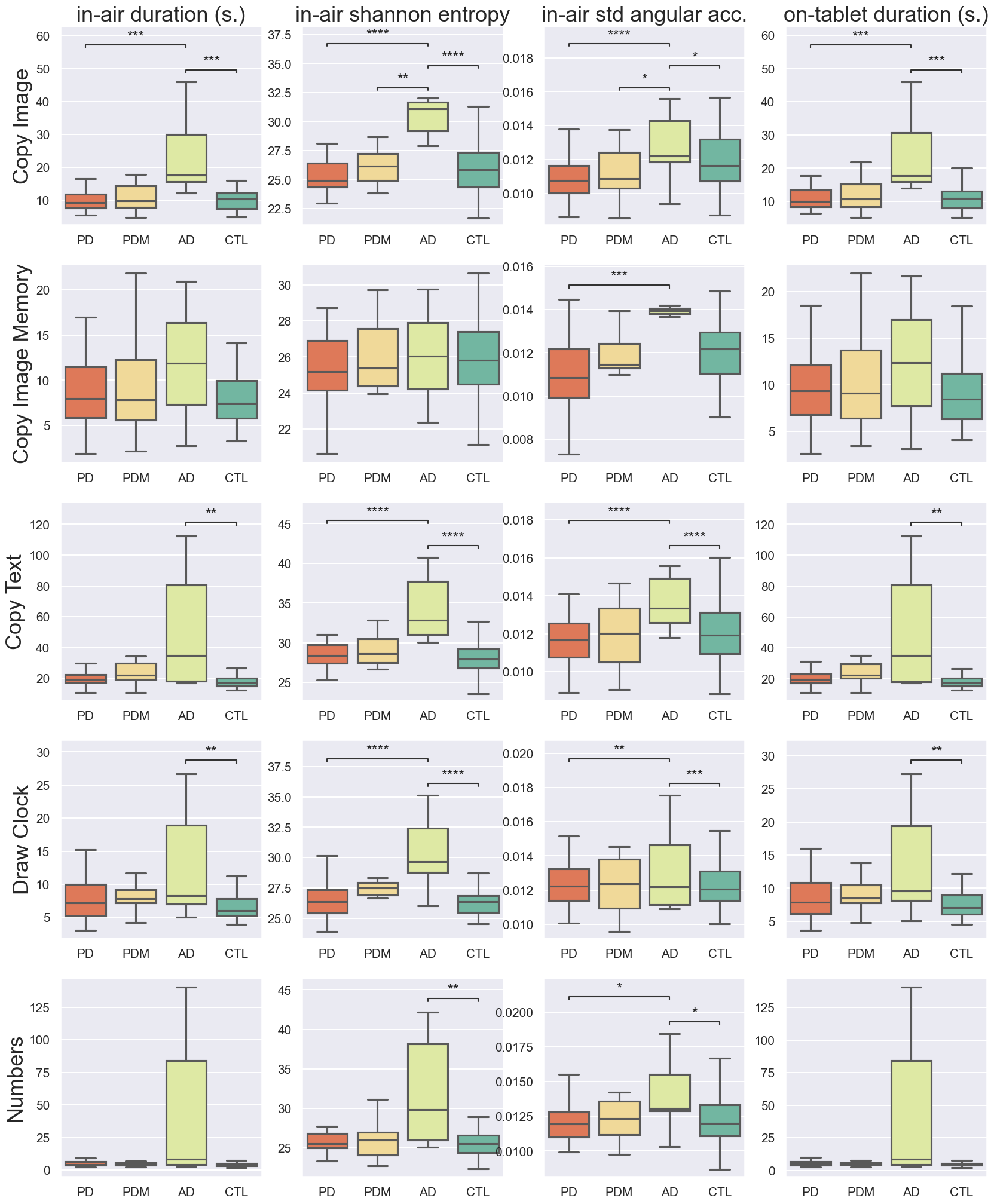}
    \caption{Boxplots of the most significant features (AD* vs CTL) presented across drawing and writing tasks for all groups of participants.}
    \label{fig:global_best_AD}
\end{figure}

Over the 54 task-agnostic features, 50 are significant ($p\leq0.05$) in comparing AD* and CTL participants for at least one task, and 29 have p-values smaller than 0.001. We present the 10 features having the lowest p-values in Table \ref{tab:global_best_AD} across all tasks. 
Our most significant and consistent features across all tasks are the Shannon entropy, the duration, the mean of the velocity and accelerations for point tasks, and the variations of angular velocity and acceleration for the drawing and writing tasks. 
We show the box plots for some of those features in Figure \ref{fig:global_best_AD}.

\textit{Shannon entropy} is a measure of the uncertainty in a signal, showing the erratic aspect of one's handwriting, which is a marker of indecision for cognitively impaired participants.
The differences measured in \textit{linear velocity} and \textit{linear acceleration}, as well as \textit{duration}, all show how significantly slower AD* participants are from the control group, presumably due to processing inefficiency.
Finally, the differences in \textit{angular velocity} and \textit{angular acceleration} are indicators of the swiftness of the writing.
For detecting AD, the most significant features were all for writing and drawing tasks, which are much more demanding cognitively than the rest of tasks and mainly consisting of tasks designed for AD* assessment, as opposed to the point and spiral tasks.

When comparing the tasks \textbf{Copy Text} and \textbf{Copy Read Text}, the difference in velocity and horizontal acceleration is significantly higher for AD* participants than CTL participants ($p$-value $\leq$ 0.001 for both), which shows how the multi-tasking has a way higher effect on cognitive participants.

A limitation of our approach here is the lower number of participant for the AD/MCI category, which did not allowed us a separated analysis of the cognitive impairments against the participants presenting only AD's signs.

\begin{table}[t]
 
    \centering
    \caption{$p$-values and AUC for the best 10 features for AD* vs CTL comparison. The table is repeated to fit the page. Significant $p$-values are \textbf{bolded}, as well as AUC values over 0.75}
    \resizebox{\linewidth}{!}{
    \begin{tabular}{l c c c c c c c c}
	\toprule

AD* vs CTL & \multicolumn{2}{c}{Copy Image}	& \multicolumn{2}{c}{Copy Text}	& \multicolumn{2}{c}{Copy Read Text}	\\
& p-val & AUC	& p-val & AUC	& p-val & AUC	\\ 
	\midrule
in-air avg. angular acc.	& 0.070& 0.700& \textbf{$<$0.001}& \textbf{0.791}& 0.087& 0.672\\
in-air shannon entropy	& \textbf{$<$0.001}& \textbf{0.873}& \textbf{$<$0.001}& \textbf{0.887}& \textbf{$<$0.001}& \textbf{0.827}\\
in-air std angular acc.	& \textbf{0.019}& 0.737& \textbf{$<$0.001}& \textbf{0.818}& \textbf{0.034}& 0.702\\
in-air std angular velocity	& \textbf{0.019}& 0.730& \textbf{$<$0.001}& \textbf{0.826}& 0.058& 0.692\\
on-tablet avg. X acc.	& \textbf{0.017}& 0.732& 1.000& 0.605& \textbf{$<$0.001}& \textbf{0.761}\\
on-tablet avg. Y acc.	& 0.078& 0.680& 1.000& 0.597& \textbf{$<$0.001}& \textbf{0.754}\\
	\midrule

AD* vs CTL & \multicolumn{2}{c}{Numbers}	& \multicolumn{2}{c}{Draw Clock}	& \multicolumn{2}{c}{Freewrite}	\\
& p-val & AUC	& p-val & AUC	& p-val & AUC	\\ 
	\midrule
in-air avg. angular acc.	& 0.125& 0.691& \textbf{$<$0.001}& \textbf{0.761}& \textbf{$<$0.001}& \textbf{0.760}\\
in-air shannon entropy	& \textbf{$<$0.001}& \textbf{0.794}& \textbf{$<$0.001}& \textbf{0.866}& \textbf{$<$0.001}& \textbf{0.769}\\
in-air std angular acc.	& \textbf{0.045}& 0.727& \textbf{$<$0.001}& \textbf{0.768}& \textbf{$<$0.001}& \textbf{0.821}\\
in-air std angular velocity	& 0.125& 0.694& \textbf{$<$0.001}& \textbf{0.799}& \textbf{$<$0.001}& \textbf{0.774}\\
on-tablet avg. X acc.	& 1.000& 0.535& \textbf{$<$0.001}& \textbf{0.788}& \textbf{0.023}& 0.698\\
on-tablet avg. Y acc.	& 1.000& 0.597& \textbf{$<$0.001}& \textbf{0.780}& \textbf{0.023}& 0.700\\
	\midrule

AD* vs CTL & \multicolumn{2}{c}{Point Dominant}	& \multicolumn{2}{c}{Point Non Dominant}	& \multicolumn{2}{c}{Point Sustained}	\\
& p-val & AUC	& p-val & AUC	& p-val & AUC	\\ 
	\midrule
in-air avg. angular acc.	& \textbf{0.016}& 0.726& 1.000& 0.627& \textbf{0.030}& 0.740\\
in-air avg. acc.	& \textbf{$<$0.001}& \textbf{0.792}& \textbf{$<$0.001}& \textbf{0.850}& \textbf{$<$0.001}& \textbf{0.809}\\
in-air avg. velocity	& \textbf{$<$0.001}& \textbf{0.797}& \textbf{$<$0.001}& \textbf{0.857}& \textbf{$<$0.001}& \textbf{0.827}\\
in-air shannon entropy	& \textbf{$<$0.001}& \textbf{0.865}& \textbf{$<$0.001}& \textbf{0.810}& \textbf{$<$0.001}& \textbf{0.872}\\
in-air std angular acc.	& \textbf{$<$0.001}& \textbf{0.804}& 0.117& 0.708& \textbf{0.003}& \textbf{0.789}\\
in-air std acc.	& \textbf{0.015}& 0.716& \textbf{$<$0.001}& \textbf{0.766}& 0.290& 0.655\\
in-air std velocity	& 0.159& 0.650& \textbf{0.004}& 0.737& 0.390& 0.647\\
\bottomrule
        \end{tabular}}
    \label{tab:global_best_AD}
     
\end{table}

\subsubsection{PD vs CTL}
\begin{figure}[ht]
    \centering
    \includegraphics[width=\linewidth]{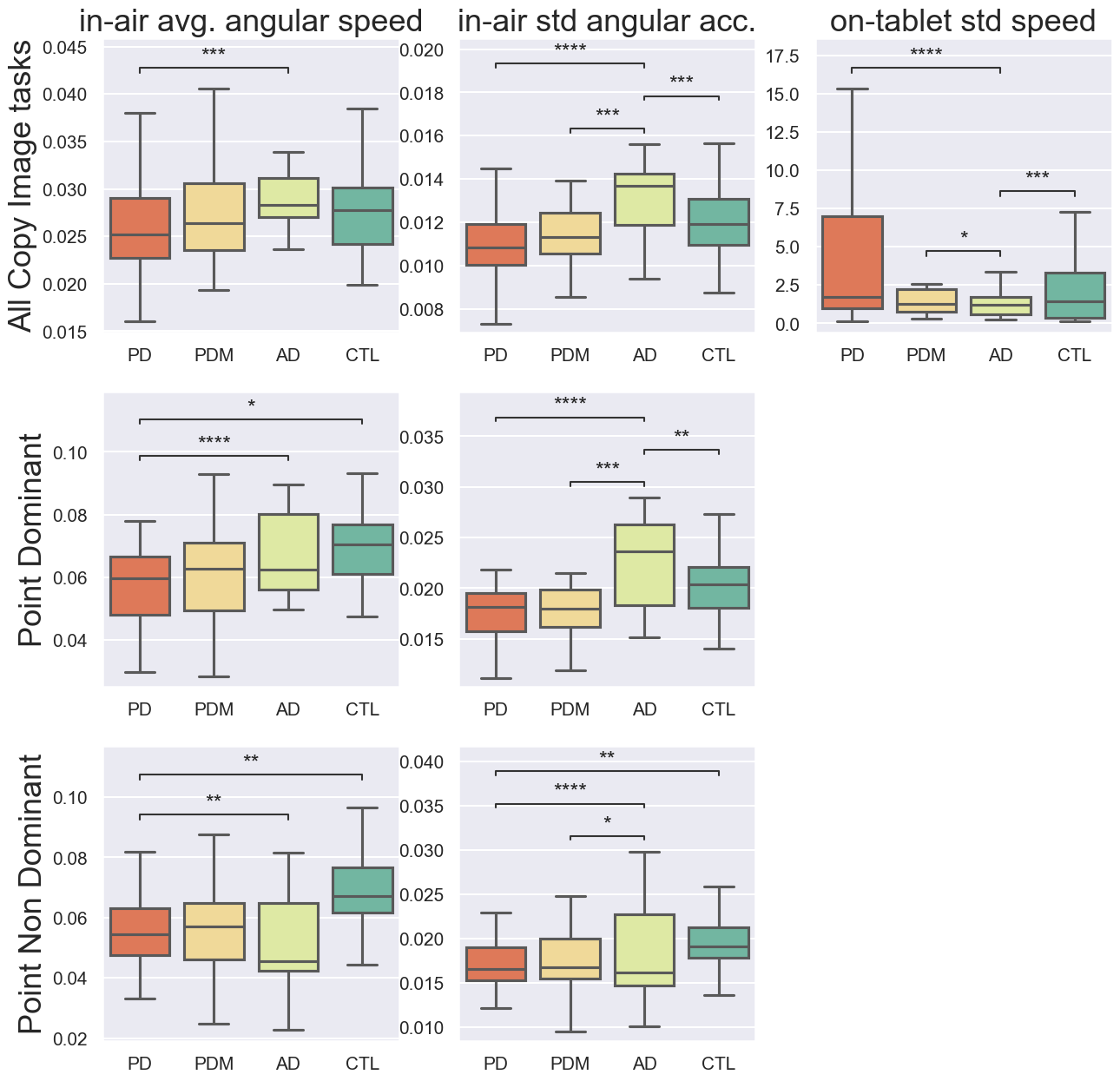}
    \caption{Boxplots of some the best features extracted for PD vs CTL, all from angular velocity and accelerations, presented for drawing and point tasks for all groups of participants.}
    \label{fig:global_best_PD}
     
\end{figure}

Over the 54 task-agnostic features, 10 are significant ($p\leq0.05$) in the comparison between PD and CTL participants for at least one task, and 7 have p-values under 0.001.
Table \ref{tab:global_best_PD} shows the top 10 features with the lowest p-values across all tasks for the PD vs. CTL comparison.
\begin{table}[ht]
    \centering
        \caption{$p$-values and AUC for the best 10 features for PD vs CTL comparison. The table is repeated to fit the page. Significant p-values are \textbf{bolded}, as well as AUC values over 0.75.}
    \resizebox{\linewidth}{!}{
    \begin{tabular}{l c c c c c c c c c c}
\toprule
PD vs CTL & \multicolumn{2}{c}{All Points}	& \multicolumn{2}{c}{Point Dominant}	& \multicolumn{2}{c}{Point Non Dominant}	\\
& p-val & AUC	& p-val & AUC	& p-val & AUC	\\ 
	\midrule
in-air avg. angular acc.	& \textbf{$<$0.001}& 0.723& \textbf{0.041}& 0.723& \textbf{0.009}& \textbf{0.768}\\
in-air avg. acc.	& \textbf{0.027}& 0.611& 1.000& 0.611& 0.443& 0.624\\
in-air avg. X acc.	& \textbf{$<$0.001}& 0.637& 0.659& 0.628& \textbf{0.032}& 0.694\\
in-air avg. Y acc.	& \textbf{$<$0.001}& 0.650& 0.076& 0.676& 0.322& 0.647\\
in-air avg. angular velocity	& \textbf{$<$0.001}& 0.742& \textbf{0.021}& 0.743& \textbf{0.008}& \textbf{0.768}\\
in-air avg. velocity	& \textbf{0.031}& 0.608& 1.000& 0.620& 0.437& 0.630\\
in-air avg. X velocity	& \textbf{$<$0.001}& 0.639& 1.000& 0.616& 0.129& 0.678\\
in-air avg. Y velocity	& \textbf{$<$0.001}& 0.682& \textbf{0.041}& 0.711& 0.143& 0.678\\
in-air std angular acc.	& \textbf{$<$0.001}& 0.723& 0.052& 0.738& \textbf{0.010}& \textbf{0.770}\\
in-air std angular velocity	& \textbf{$<$0.001}& 0.733& \textbf{0.021}& \textbf{0.756}& \textbf{0.009}& \textbf{0.759}\\

\bottomrule

    \end{tabular}}

    \label{tab:global_best_PD}
\end{table}
These features primarily revolve around in-air dynamics, variations, and averages of velocitys and angular velocitys. They are notably discriminatory for point tasks. Additionally, we present box plots for these significant features in Figure \ref{fig:global_best_PD} for further visualization and analysis.

For detecting PD, the most significant features were for drawing and point tasks. 
The point tasks were better suited to assess motoric features of PD, such as tremor and bradykinesia. Accordingly, we observed subjects with PD showed significantly lower \textit{angular velocity} and \textit{angular acceleration} in air, which could be indicators of hypometria \cite{broderick2009hypometria}. 
The drawing tasks were much more complex tasks, mainly designed for cognitive evaluation but that still require basic motor skills to complete.
In those tasks, the \textit{in-air angular velocity} is still a discriminative feature, but features of variation (as standard deviation or amplitudes) of the velocity and acceleration on the tablet were notable. These findings may represent, tremor, dystonia, dyskinesia, myoclonus other other complicating motor symptoms that develop in some persons with PD~\cite{smits2014standardized}.
However, none of the task-agnostic features measured on the spirals tasks, mainly motor tasks often used for PD assessment, were discriminative enough.

\subsubsection{PD vs PDM}
Among the 54 task-agnostic features, only 6 are significant ($p\leq0.05$) in comparing PD and PDM participants for at least one task. We only found significant features for the \textbf{Spiral} tasks here.

\begin{figure}[ht]
    \centering
    \includegraphics[width=\linewidth]{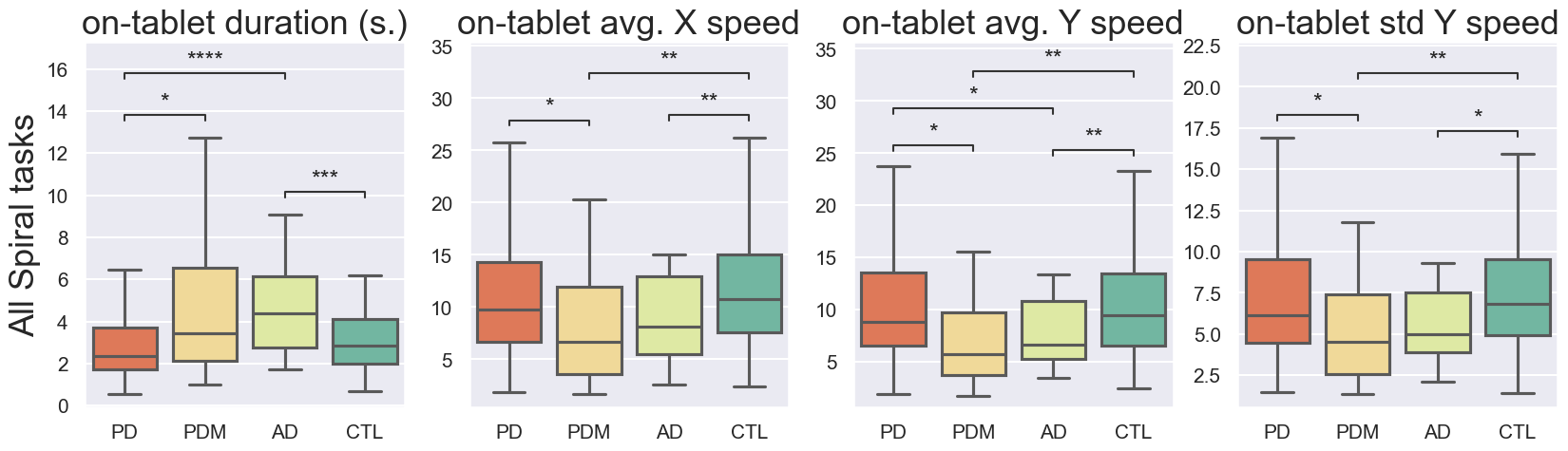}
    \caption{Boxplots of some the best features extracted for PD vs PDM for all the spiral tasks combined.}
    \label{fig:PD_PDM_spiral}
     
\end{figure}

To distinguish PD from other parkinsonism, significant variations can only be found in the \textbf{Spiral} tasks, motor exercises designed for PD assessment.
We observe few significant features for the \textbf{Spiral} tasks, mostly amplitudes of \textit{velocity} and \textit{acceleration}, the best being for the \textit{vertical velocity}.
The \textbf{Spiral} tasks reveal a difference in velocity between PD and PDM participants, the latest being slower, as shown in Figure \ref{fig:PD_PDM_spiral}.

\begin{figure}[ht]
    \centering
    \includegraphics[width=\linewidth]{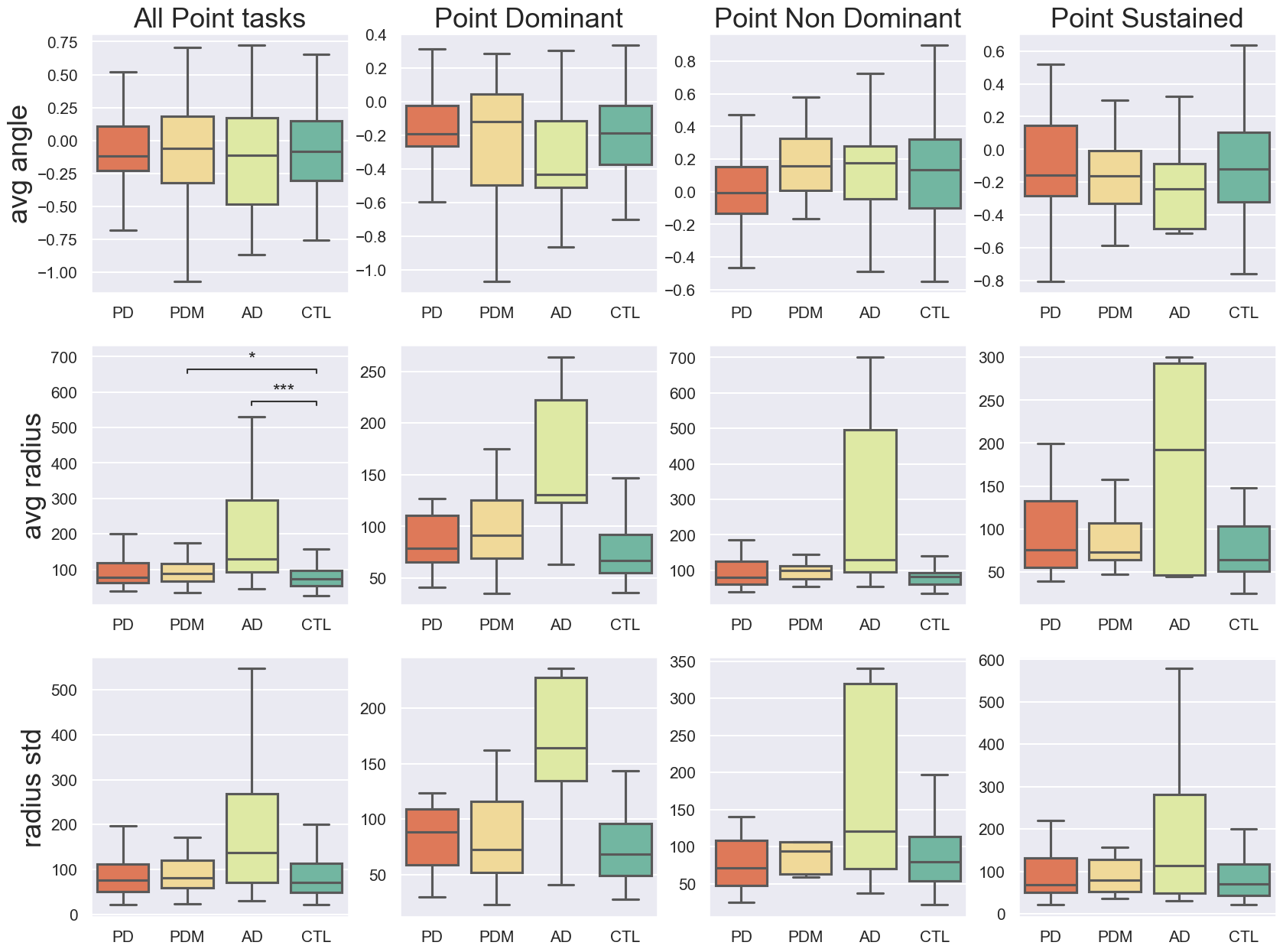}
    \caption{Boxplots of the best point-specific features for all groups of participants.}
    \label{fig:specific_points}
\end{figure}

\vspace{-4mm}
\subsection{Task-Specific Results}
\label{ssec:specific_res}
After presenting the results of the task-agnostic features, we focus on the task-specific ones, which are crafted especially for certain tasks or groups of tasks.

\subsubsection{Point Tasks Results}

Figure \ref{fig:specific_points} shows all features with their significance.
None of the features are consistently significant across all tasks, but the \textit{radius average} and \textit{radius standard deviation} features exhibit the same behavior across all tasks: features from AD* participants are higher than in the other groups.
We observe a slightly higher \textit{radius average} for PD participants compared to CTL, only significant when considering all point tasks, showing that the \textit{Point} tasks capture only little of the expected tremors.

\subsubsection{Spiral Tasks Results} 

\begin{figure}[ht]
    \centering
    \includegraphics[width=\linewidth]{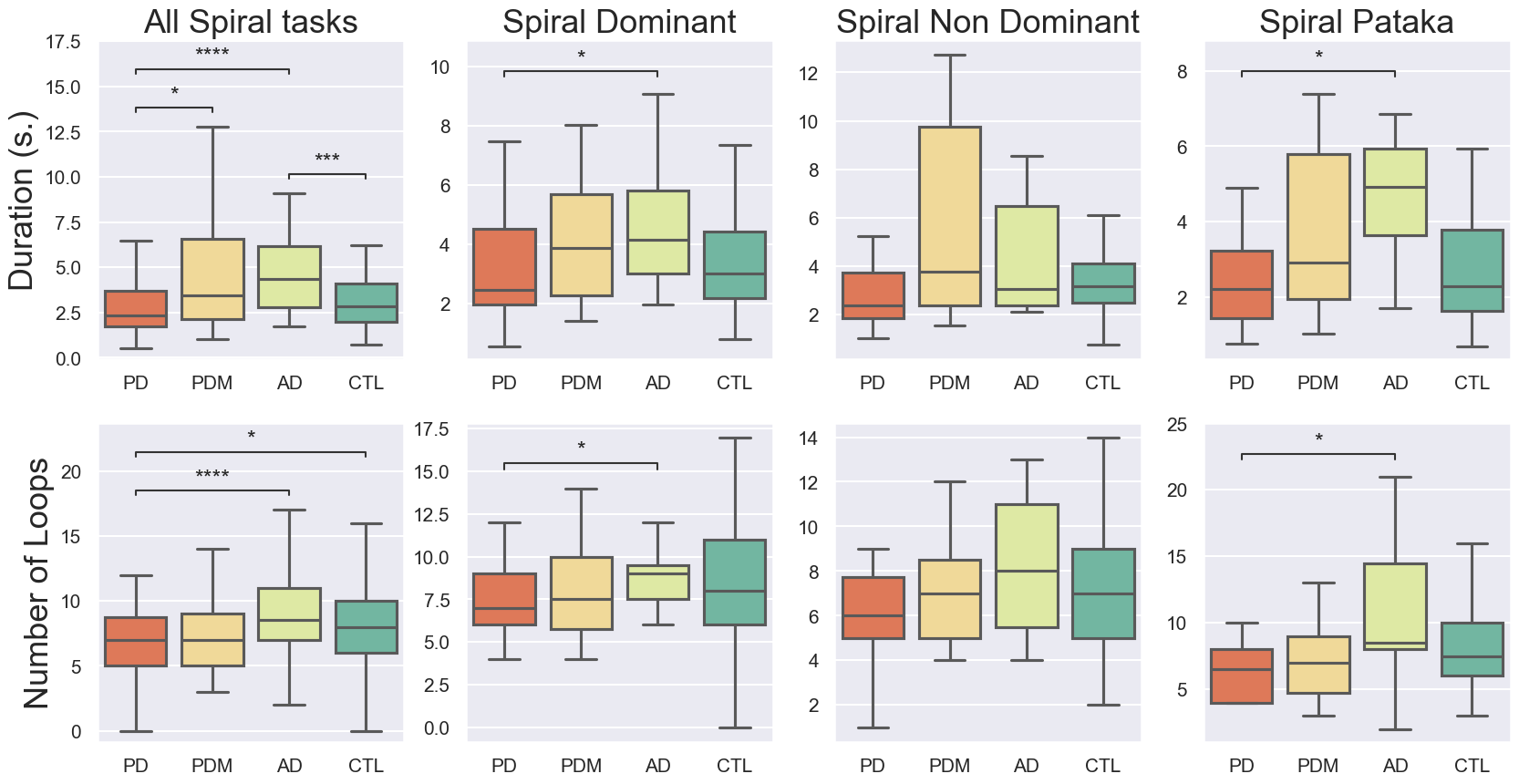}
    \caption{Boxplots of the best spiral-specific features, for all groups of participants.}
     
    \label{fig:specific_spirals}
\end{figure}

Figure \ref{fig:specific_spirals} shows the boxplots of the spiral-specific features that were significant for at least one of the tasks.
All the features exhibit a similar behavior across all tasks:
PDM and AD* participants take more time to draw their spirals than PD and CTL, AD* and CTL participants draw more loops, which gives a lower time per loop (\textit{Fundamental period of the spiral}) for PDM and AD* participants.
Subjects were not instructed to provide a minimum number of loops, therefore interpretation of task duration is difficult. The reduced number of loops in the PD cohort may have been an energetically rationale choice as the task is particularly difficult with persons with impaired dexterity. As PDM represents a mixed disease cohort, few conclusions relative to PD can be drawn. It is doubtful differences are due to cognitive impairment as overall MOCA performance was similar between these groups.

\begin{table}[t]
 
    \centering
    \caption{$p$-values and AUC of the specific features for all spiral tasks significant for at least one comparison. Significant p-values and AUC over 0.75 are \textbf{bolded}.}
    \resizebox{\linewidth}{!}{
    \begin{tabular}{l l|c c c c c c c}
	\toprule

Spirals & Feature  & \multicolumn{2}{c}{AD* vs CTL}	& \multicolumn{2}{c}{CTL vs PD}	& \multicolumn{2}{c}{PD vs PDM}	\\ 
	&	& p-val & AUC	& p-val & AUC	& p-val & AUC	\\
	\midrule
\multirow{5}{*}{Spirals}
& N loops	& 0.092& 0.573& \textbf{$<$0.001}& 0.604& 0.303& 0.538\\
& avg radius diff	& 0.141& 0.557& \textbf{0.019}& 0.568& 0.945& 0.554\\
& Duration	& \textbf{$<$0.001}& 0.695& 0.104& 0.543& \textbf{0.003}& 0.623\\
& Number of Loops	& 0.370& 0.563& \textbf{$<$0.001}& 0.615& 0.209& 0.542\\
& Fundamental Period	& \textbf{$<$0.001}& 0.694& 0.109& 0.545& \textbf{0.004}& 0.615\\
	\midrule
\multirow{2}{*}{Dominant}
& Duration	& \textbf{$<$0.001}& 0.695& 0.496& 0.543& 0.155& 0.623\\
& Fundamental Period	& \textbf{$<$0.001}& 0.694& 0.474& 0.545& 0.185& 0.615\\
	\midrule
\multirow{4}{*}{Non Dominant}
& N loops	& 0.714& 0.573& \textbf{0.018}& 0.604& 0.166& 0.538\\
& Duration	& 0.123& 0.695& \textbf{0.043}& 0.543& \textbf{0.028}& 0.623\\
& Number of Loops	& 0.674& 0.563& \textbf{0.031}& 0.615& 0.241& 0.542\\
& Fundamental Period	& 0.100& 0.694& 0.051& 0.545& \textbf{0.035}& 0.615\\
	\midrule
\multirow{4}{*}{Pataka}
& N loops	& 0.150& 0.573& \textbf{0.012}& 0.604& 0.429& 0.538\\
& Duration	& \textbf{$<$0.001}& 0.695& 0.700& 0.543& 0.105& 0.623\\
& Number of Loops	& 0.406& 0.563& \textbf{0.007}& 0.615& 0.620& 0.542\\
& Fundamental Period	& \textbf{$<$0.001}& 0.694& 0.658& 0.545& 0.103& 0.615\\
\bottomrule
    \end{tabular}}
    \label{tab:local_spiral}
     
\end{table}

\subsubsection{Numbers Task Results}

\begin{table}[ht]
    \centering
    \caption{$p$-values and AUC of the significant features specific to the numbers task. Significant p-values and AUC over 0.75 are \textbf{bolded}.}
    \resizebox{\linewidth}{!}{
    \begin{tabular}{l|c c c c c c}
	\toprule
Feature  & \multicolumn{2}{c}{AD* vs CTL}	& \multicolumn{2}{c}{CTL vs PD}	& \multicolumn{2}{c}{PD vs PDM}	\\ 
& p-val & AUC	& p-val & AUC	& p-val & AUC	\\
	\midrule

time in in-air	& \textbf{0.003}& 0.690& 0.073& 0.612& 0.978& 0.502\\
avg ans time	& \textbf{0.033}& 0.636& 0.340& 0.560& 0.366& 0.577\\
change in ans width	& 0.203& 0.581& \textbf{0.021}& 0.644& 0.116& 0.634\\
change in ans pressure	& \textbf{0.030}& 0.639& 0.577& 0.535& 0.596& 0.545\\
change in time btwn ans	& \textbf{0.007}& 0.673& 0.088& 0.607& 0.963& 0.504\\

\bottomrule
    \end{tabular}}
    \label{tab:local_numbers}

\end{table}

Table \ref{tab:local_numbers} shows the Numbers-specific features' results, studying mostly the variations in time and dimensions between successive answers in the \textbf{Numbers Task}.
Of the few features that show a significant difference between groups of participants, the time taken to answer is the first feature differentiating between AD* and CTL participants, AD* participants being slower.
In a similar fashion, AD* participants are significantly slower over time as compared to CTL participants, an effect of the increased fatigue over the various answers.
Finally, we observe that both AD* and PD participants tend to input more and more pressure over time compared to CTL participants, once again a potential fatigue indicator.

\subsubsection{Writing Tasks Results}

\begin{figure}[ht]
    \centering
    \includegraphics[width=\linewidth]{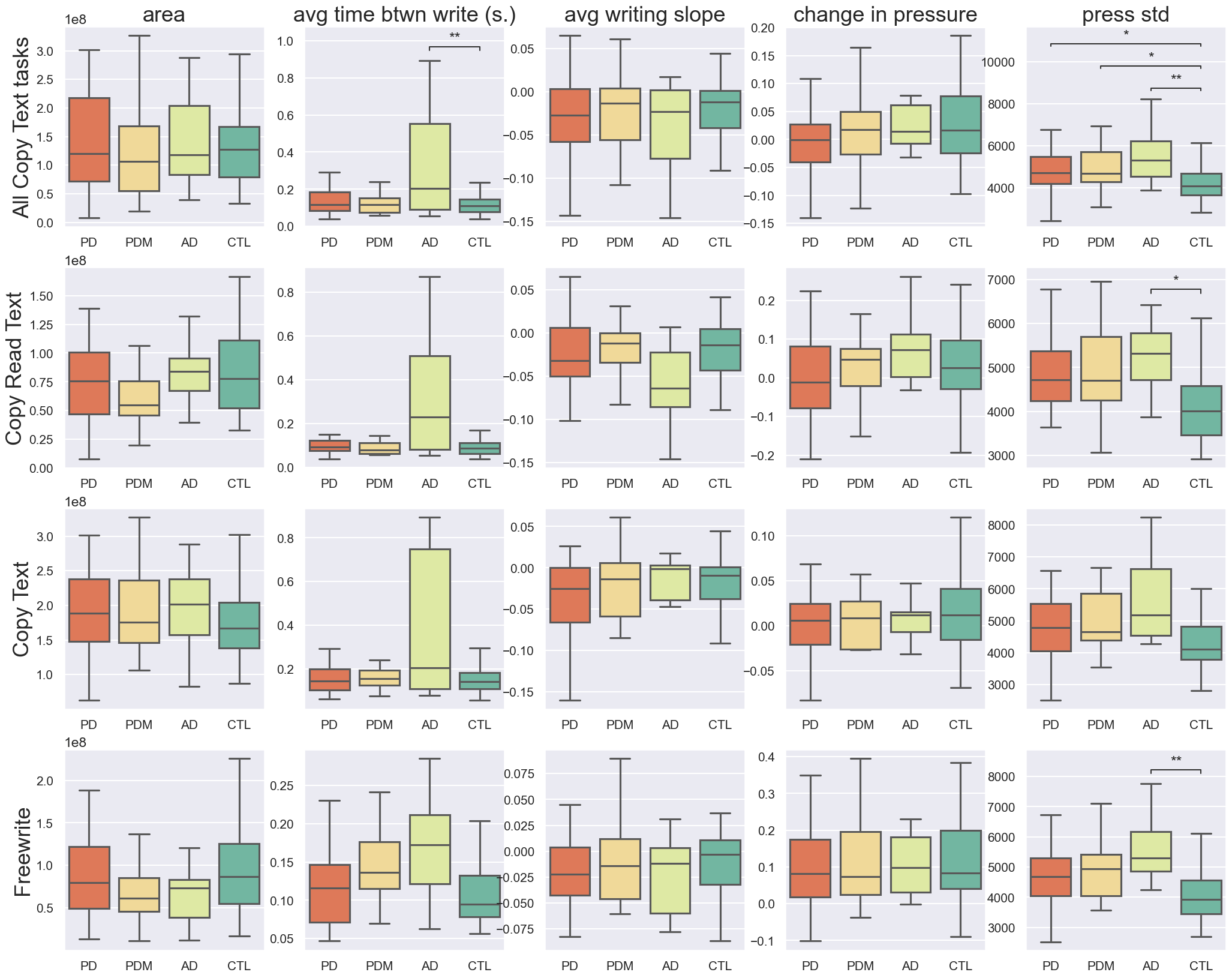}
    \caption{Boxplots of the best writing specific features for all groups of participants.}
    \label{fig:specific_write}
     
\end{figure}

In this section, we present the results for the writing tasks's specific features.
As shown in Figure \ref{fig:specific_write}, AD* participants take a much longer \textit{Average Time Between Writing} than all other participants across all tasks, revealing a longer hesitation and intermitent pausing times, between words.
CTL participants generally have lower variations in the pressure they use than ND participants, across all tasks, a potential indicator of more stability in steadiness in the writing.
PD participants use less and less pressure as they write, in contrast to the AD* and CTL participants, and they also use a larger area to write, significant only in the \textbf{Copy Text Task}, which correlates with previous findings about PD participants writing larger than others~\cite{thomas2017handwriting}.

\begin{table}[ht]
    \centering
    \caption{$p$-values and AUC of the specific features for all significant writing tasks. Significant p-values and AUC over 0.75 are \textbf{bolded}.}
    \resizebox{\linewidth}{!}{
    \begin{tabular}{l l|c c c c c c c}
	\toprule
Writing & Feature  & \multicolumn{2}{c}{AD* vs CTL}	& \multicolumn{2}{c}{CTL vs PD}	& \multicolumn{2}{c}{PD vs PDM}	\\ 
	&	& p-val & AUC	& p-val & AUC	& p-val & AUC	\\
	\midrule
\multirow{2}{*}{Copy Read Text}
& press std	& \textbf{0.004}& 0.729& \textbf{$<$0.001}& 0.700& 0.602& 0.542\\
& avg time btwn write	& \textbf{0.012}& 0.639& 0.303& 0.539& 0.477& 0.674\\
	\midrule
\multirow{3}{*}{Copy Text}
& press std	& \textbf{0.027}& 0.729& \textbf{0.017}& 0.700& 0.508& 0.542\\
& avg time btwn write	& \textbf{0.004}& 0.639& 0.835& 0.539& 0.495& 0.674\\
& avg x dist btwn write	& \textbf{0.005}& 0.617& 0.459& 0.507& 0.967& 0.647\\
	\midrule
\multirow{2}{*}{Freewrite}
& press std	& \textbf{$<$0.001}& 0.729& \textbf{$<$0.001}& 0.700& 0.628& 0.542\\
& avg time btwn write	& \textbf{0.019}& 0.639& 0.539& 0.539& \textbf{0.046}& 0.674\\
\bottomrule
    \end{tabular}}
    \label{tab:local_writing}
     
\end{table}

\vspace{-4mm}
\subsection{Classification Results}

This subsection discusses the classification results obtained from the different handwriting tasks analyzed in this study. The results are discussed by examining each group comparison individually. The models' performance is evaluated in terms of accuracy (ACC), F-1 score (F1), and AUC. Table \ref{tab:classification_results} presents results obtained using the Bagging classifier. Since the BG classifier consistently outperformed the other two classifiers we tested—RF and MLP—in terms of accuracy, F1-score, and AUROC in the different tasks considered, we present and discuss the results obtained with this classifier in the main text for the sake of simplicity. The experimental results for RF and MLP are provided in the Appendix.

\begin{table*}[ht]
      \caption{Classification results for pairwise group comparisons, using the Bagging model. The table reports experimental results in terms of accuracy (ACC), F1-score (F1), and AUC for each comparison. The best results for each feature for each group comparison are indicated in bold, while the second-best are italicized.}
\Large
\label{tab:classification_results}
\center
    \resizebox{0.87\textwidth}{!}{
\begin{tabular}{l|ccc|ccc|ccc|ccc|ccc|ccc}
\toprule
\multirow{2}{*}{\textbf{Task}} & \multicolumn{3}{c|}{\textbf{AD* vs PD}} & \multicolumn{3}{c|}{\textbf{PD vs PDM}} & \multicolumn{3}{c|}{\textbf{AD* vs CTL}} & \multicolumn{3}{c|}{\textbf{AD* vs PDM}} & \multicolumn{3}{c|}{\textbf{PD vs CTL}} & \multicolumn{3}{c}{\textbf{PDM vs CTL}} \\ 
 & ACC & F1 & AUC & ACC & F1 & AUC & ACC & F1 & AUC & ACC & F1 & AUC & ACC & F1 & AUC & ACC & F1 & AUC \\

 \midrule

Copy Cube & 71 & 68 & 81 & 50 & 48 & 57 & 67 & 66 & 69 & 67 & 64 & 69 & 64 & 63 & \textbf{71} & 57 & 51 & 55 \\ 
Draw Clock & 86 & 86 & 90 & 56 & 55 & 61 & \textbf{87} & \textbf{87} & \textit{88} & 79 & 77 & 71 & 60 & 60 & 54 & 67 & 53 & 39 \\ 
Copy Image & \textbf{91} & \textbf{91} & 91 & \textit{67} & 53 & 46 & 82 & 82 & 86 & 78 & 77 & \textit{81} & 48 & 48 & 54 & 56 & 52 & 35 \\ 
Copy Image Memory & 73 & 73 & 83 & 65 & 51 & 51 & 64 & 63 & 62 & 67 & 67 & 63 & \textbf{69} & \textbf{69} & 69 & 54 & 48 & 47 \\ 
\midrule
Copy Read Text & 77 & 75 & 81 & 59 & 44 & 35 & 67 & 67 & 72 & 76 & 74 & 78 & 61 & 60 & 62 & \textit{71} & 69 & 79 \\ 
Copy Text & 77 & 78 & 88 & 63 & 58 & \textbf{74} & 81 & 82 & \textbf{89} & 73 & 72 & \textbf{83} & 63 & 63 & 64 & 67 & 65 & 66 \\ 
Freewrite & 69 & 69 & 76 & 66 & 52 & 45 & 71 & 71 & 75 & 73 & 70 & 65 & 48 & 48 & 47 & \textbf{77} & \textbf{77} & \textbf{83} \\ 
Numbers & 69 & 69 & 72 & 57 & 49 & 45 & 68 & 68 & 75 & 63 & 57 & 53 & 55 & 55 & 53 & 65 & 51 & 42 \\ 
\midrule
Point Dominant & \textbf{91} & \textbf{91} & \textit{93} & 66 & 52 & 47 & 84 & \textit{84} & 85 & \textit{82} & \textit{82} & 78 & 66 & 65 & \textit{68} & 57 & 55 & 53 \\ 
Point Non-Dominant & 84 & 84 & \textbf{94} & \textit{67} & 56 & 52 & \textit{85} & \textit{84} & 85 & 73 & 71 & 72 & \textit{68} & \textit{68} & 66 & \textit{71} & 71 & \textit{72} \\ 
Point Sustained & \textit{89} & \textit{89} & 92 & \textit{67} & 54 & 59 & 80 & 80 & \textbf{89} & \textbf{83} & \textbf{83} & 79 & 62 & 62 & 69 & 65 & 65 & 59 \\ 
\midrule
Spiral Dominant & 67 & 64 & 69 & \textit{67} & 53 & 50 & 60 & 60 & 63 & 75 & 69 & 57 & 54 & 52 & 54 & 60 & 57 & 60 \\ 
Spiral Non-Dominant & 65 & 63 & 59 & 64 & \textit{63} & \textit{66} & 57 & 55 & 58 & 79 & 78 & 64 & 41 & 41 & 33 & 60 & 57 & 50 \\ 
Spiral PaTaKa & 70 & 68 & 71 & \textbf{73} & \textbf{71} & 51 & 64 & 61 & 56 & 73 & 68 & 68 & 60 & 60 & 61 & \textbf{77} & \textit{76} & 69 \\ 

\bottomrule
\end{tabular}
}
 
\end{table*}

When comparing AD* against PD, the \textbf{Copy Image} and the \textbf{Point Dominant} tasks exhibit the best performance in terms of accuracy (91) and F1-Score (91). The \textbf{Copy Image} task involves visual perception and fine motor skills, the latter of which is particularly impaired in PD. Moreover, the \textbf{Point Non-Dominant} task reports the highest AUC score (94), indicating its efficacy in discriminating between the two conditions \cite{jose2020altered}. This is intriguing as mirror movement or mirroring dystonia have been described in Task Specific Dystonia and other movement disorders as a way to unmask and characterize latent dystonia features~\cite{quattrone2023reflection}; in our PD cohort, this task may be abnormal in PD subjects who are prone to the development of dystonic features - a finding generally thought to be absent in AD.

In the comparison between the PD and PDM group, the \textbf{Spiral PaTaKa} task stands out with the highest accuracy (73) and F1 score (71). Spiral drawing tasks are known for their sensitivity in assessing motor skills due to their complexity and demand for coordinated movements. Given the significant impact of PD on motor function, tasks like spiral drawing can provide valuable insights into motor impairment severity. The task reporting the highest AUC is \textbf{Copy Text}.

For the comparison between AD* and CTL, the \textbf{Draw Clock} task achieves the highest accuracy (87) and F1 score (87). This task assesses the ability to draw a clock with two hands correctly placed. This complex task includes a wide set of skills, including adequate function of spatial awareness, executive planning, error detection, motor coordination and execution as well as language processing and numerical processing. This task is often used to screen for intra-cortical and subcortical brain network deficiencies, observed notably in individuals with early-stage dementia. The high discriminability of this task could be attributed to cognitive impairment present in the AD* group, as indicated by their lower MoCA score compared to other groups. Additionally, the \textbf{Copy Text} and \textbf{Point Sustained} tasks yield the highest AUC, indicating their utility in distinguishing between AD* and CTL groups.

When contrasting AD* with PDM, the highest accuracy (83) and the F1 score (83) are achieved in the \textbf{Point Sustained} task. Given the attentional deficits commonly observed in AD, this task's effectiveness in discriminating between AD* and CTL and PDM groups is plausible.

In comparing PD and CTL groups, the \textbf{Copy Image Memory} task displays the best performance with an accuracy and F1 score of 69. This task assesses memory function, executive function, and visuospatial skills, which are often impaired in individuals with PD compared to CTLs. Additionally, the \textbf{Copy Cube} task yields the highest AUC, indicating its effectiveness in this comparison.

Finally, when comparing the CTL and PDM groups, the \textbf{Freewrite} task emerges as the most effective, achieving the highest accuracy and F1 score of 77 and an AUC of 83. This task involves language production and cognitive flexibility, domains commonly affected in individuals with PDM compared to healthy controls. The cognitive impairment exhibited by the PDM group is further confirmed by their low MoCA score (see Table X). 

\vspace{-4mm}
\subsection{Correlation of the features with UPDRS-III and MoCa scores}
To further explore the clinical relevance of the extracted handwriting features, we measure the correlation between each of the 54 task-agnostic features and the MoCA score~\cite{julayanont2017montreal}, which evaluates global cognitive functioning, and the motor section of the Unified Parkinson's Disease Rating Scale (UPDRS-III)~\cite{movement2003unified}, as well as each of the 14 subdomains of the UPDRS-III.

First we verified the normality of the distribution of the scores, and the distribution of features for each task, by using a Shapiro-Wilk test.
All scores passed the normality test (p-values$\leq 10^{-18}$) and all features passed the normality test (p-values$\leq 0.037$).

Then we verify the linearity condition between our features and the scores previously mentioned by fitting a linear and a quadratic function to each pair of features/scores, and perform a Nested model F-test over the residual sums of squares of both functions, to find out if adding the quadratic component makes a significant difference. Most features did not pass the linearity test with the various scores (3.4\% to 16.3\% had a p-value under 0.05, depending on the score at hand).
As most features do not pass the linearity test, we use the Spearman correlation to measure the correlation between each score and each feature.

None of the features were significantly correlated with UPDRS-III scores ($p < 0.05$), suggesting limited association between our handwriting features and motor symptom severity as assessed by UPDRS-III. 
However, UPDRS-III is a conglomerate of many sub-domains, for which some are not related to handwriting at all.
A categorical examination of each subdomain of the UPDRS-III revealed a great variation of correlations depending on the domains, as show in Table \ref{tab:updrs_corr}. The complete list of features with their individual significance is available in the appendix.

\begin{table}[ht]
    \centering
    \caption{Number of features over the 54 task-agnostic features produced being correlated with MoCa score, UPDRS-III score and its sub-components.}
    \resizebox{\linewidth}{!}{
    \begin{tabular}{l c c c c}
	\toprule
    \multirow{2}{*}{Score} & \multicolumn{4}{c}{$p\_value<$}\\
 & $0.05$ & $0.01$ & $10^{-3}$ & $10^{-4}$ \\
\midrule
\midrule
MoCa~\cite{julayanont2017montreal} & 30	& 18	& 11	& 9 \\ 
\midrule
UPDRS-III~\cite{movement2003unified} & 0	& 0	& 0	& 0 \\ 
\midrule
Rigidity & 0	& 0	& 0	& 0 \\ 
Upper bradykinesia & 2	& 2	& 0	& 0 \\ 
Lower bradykinesia & 28	& 17	& 2	& 0 \\ 
Arising from chair & 15	& 4	& 1	& 0 \\ 
Gait & 19	& 0	& 0	& 0 \\ 
Freeze of gait & 16	& 6	& 0	& 0 \\ 
Postural stability & 7	& 3	& 0	& 0 \\ 
Posture & 0	& 0	& 0	& 0 \\ 
Gait and posture composite & 0	& 0	& 0	& 0 \\ 
Kinetic tremor & 22	& 0	& 0	& 0 \\ 
Postural Tremor & 6	& 2	& 0	& 0 \\ 
Resting tremor & 6	& 0	& 0	& 0 \\ 
Global tremor & 33	& 3	& 0	& 0 \\ 
Non-tremor components & 1	& 0	& 0	& 0 \\ 
    \bottomrule
    \end{tabular}}
    \label{tab:updrs_corr}
\end{table}

In contrast, a substantial number of features (30 out of 54) were significantly correlated with MoCA scores ($p < 0.05$), indicating a strong relationship between handwriting behavior and cognitive performance. Among these, 11 features exhibited highly significant correlations with $p$-values below 0.0001, underscoring the potential of handwriting-based features as sensitive indicators of cognitive impairment.

 \vspace{-4mm}
\section{Discussion and Conclusion}
\label{sec:conclusion}
The primary goal of this study was to explore the utility of handwriting as a digital biomarker for the assessment and characterization of neurodegenerative diseases (NDs), including Alzheimer's disease (AD), Parkinson’s disease (PD), and related Parkinsonian disorders (PDM). We introduced a novel dataset collected at Johns Hopkins Hospital from 113 participants (46 healthy controls and 73 ND patients) who completed 14 structured handwriting tasks on a digital tablet. This rich dataset comprises 1840 handwriting recordings and is part of a broader multimodal acquisition effort that also includes synchronized speech and eye-tracking data.

We proposed and evaluated a large array of interpretable handwriting features designed to capture motor, cognitive, and behavioral markers from handwriting signals. These features were categorized as either task-agnostic (general signal descriptors applicable across tasks) or task-specific (features tailored to the expected behavior in particular task types). Our aim was to assess the ability of these features to distinguish between ND subtypes and healthy controls using statistical analysis and binary classification experiments, while ensuring clinical interpretability and relevance.

\vspace{-4mm}
\subsection{Task-Agnostic and Task-Specific Features: Findings and Limitations}
Our analysis of task-agnostic features, those computed independently of the specific task performed, demonstrated that a substantial number (53 out of 54) significantly differentiated at least one group comparison (e.g., PD vs. control, AD* vs. PDM). These features included velocity, acceleration, angular velocity, entropy, and timing-related features. Notably, these general features were consistent across task categories and aligned with known ND symptomatology. For example, reduced velocity and increased entropy were particularly salient in AD* participants, aligning with executive dysfunction and impaired motor planning.

However, the task-agnostic approach also revealed limitations. While general features captured high-level differences, they occasionally failed to detect hallmark disease-specific signs. Most notably, micrographia, a classic symptom of PD, was not captured in a statistically significant way. This suggests that some ND-specific features may only manifest under conditions that tightly control for task type, hand dominance, or cognitive load.

To address this limitation, we introduced task-specific features, tailored to the intent and structure of each task. These revealed more nuanced and disease-relevant patterns. AD* participants showed longer in-air times, slower execution, and greater pressure variability, particularly during cognitively demanding tasks such as paragraph writing and arithmetic. PD participants drew fewer spiral loops, moved more slowly during motor tasks, and exhibited greater pressure variability, overlapping with patterns observed in the PDM group. While the PDM group was younger, its behavior often fell between the PD and control groups, suggesting that our features are relatively robust to age differences.

Despite these strengths, cross-task comparisons (such as between dominant and non-dominant hand usage or between reading and non-reading conditions) showed limited discriminative power, with the exception of Copy Text versus Copy Read Text. These results suggest the need for more advanced methods to capture inter-task behavioral adaptations.

\vspace{-4mm}
\subsection{Classification Performance and Observed Limitations}
In this study, we prioritized classification using interpretable features, incorporating task-specific and general features across handwriting tasks. Our primary goal was to develop a model that provides meaningful insights into the motor and cognitive impairments associated with neurodegenerative diseases, ensuring that the extracted features remain clinically interpretable. This approach allows for a better understanding of the underlying mechanisms driving classification decisions, which is crucial for real-world applications in clinical settings. However, deep learning-based approaches, such as those leveraging raw handwriting images or spectrogram representations of handwriting signals, offer a promising alternative. In a recent study \cite{laouedj2025detecting}, we explored non-interpretable methods by applying CNN and CNN-BLSTM models to spectrogram representations of handwriting signals for neurodegenerative disease classification. Our results demonstrated that classification performance varied across handwriting tasks and spectrogram channel combinations, with CNN consistently outperforming CNN-BLSTM. While the current study focused on interpretable features, future work could investigate hybrid approaches integrating deep learning techniques with interpretable models to enhance classification performance while maintaining clinical transparency. This would allow for the benefits of deep learning’s feature extraction capabilities while ensuring that the results remain interpretable and useful for clinical decision-making.

\vspace{-4mm}
\subsection{Clinical Applications and Implications}
Handwriting analysis has strong potential as a non-invasive, low-cost, and accessible digital biomarker. It captures both motor and cognitive processes and can be easily deployed using commercial tablets or styluses. Clinically, such tools could support early screening, aid differential diagnosis, and monitor disease progression or response to treatment. The features developed in this study are interpretable and aligned with known clinical features, making them suitable for real-world use. 

While handwriting analysis shows significant promise as a digital biomarker for neurodegenerative diseases, the current equipment setup remains a notable limitation for widespread clinical adoption. The data in this study were collected using a high-resolution graphics tablet capable of capturing position, pressure, and in-air movements with high precision. However, this hardware is relatively expensive and requires controlled conditions and trained personnel for operation. These factors restrict its applicability outside research environments and limit deployment in routine clinical settings or home-based assessments.

\vspace{-4mm}
\subsection{Discussion}
While the proposed features and classification results offer compelling evidence for the utility of handwriting in distinguishing neurodegenerative conditions, there are several limitations that affect generalizability. 
First, the models developed in this study are trained on a dataset collected under tightly controlled conditions with a homogeneous set of equipment and task protocols. Their applicability to new populations, clinical settings, or hardware setups remains to be validated.
This lack of generalization could be addressed through the use of pre-trained systems, such as neural networks which can learn more transferable representations, or through the use of more diverse handwriting datasets. 

Additionally, while handwriting captures a rich combination of motor and cognitive activity, it is only one window into neurodegeneration. The broader dataset collected in this study also includes speech and eye-tracking data, which have yet to be analyzed in combination with handwriting. Multimodal approaches are likely to improve both sensitivity and specificity by modeling interactions across domains such as language, motor planning, and attention.

Finally, extending the current dataset by increasing the number of participants, balancing demographic variables, and adding more task repetitions will improve statistical power and enable more reliable model development and evaluation.

\vspace{-4mm}
\subsection{General Conclusion and Future Directions}
This study introduced a diverse set of interpretable features for handwriting analysis and demonstrated their value in differentiating between neurodegenerative conditions using data from structured tablet-based tasks. The results validate handwriting as a meaningful digital biomarker and provide a strong foundation for future research and clinical translation.

Building on this work, our future efforts will focus on four main directions. First, we plan to increase the size and diversity of our dataset to improve model robustness and enable the study of more subtle effects, such as early-stage impairments and cross-condition overlap. Second, we will integrate the available speech and eye-tracking data to explore multimodal models that can capture interaction effects across cognitive and motor channels. Third, we will investigate neural network-based solutions, including the use of pre-trained models, to leverage broader datasets and improve generalizability across populations and devices. Finally, we will evaluate the performance of our models on more portable and affordable equipment, with the goal of developing lightweight and scalable tools suitable for deployment in real-world clinical and home settings.

Together, these future directions aim to bridge the gap between research prototypes and practical, interpretable, and accessible tools for early detection and monitoring of neurodegenerative diseases.

 \vspace{-4mm}

\bibliographystyle{IEEEtran} 
\bibliography{main}

\appendix 


\section{Additional Tables}
\begin{table*}[ht]
    \centering
    \caption{Number of files and participants for each task and each experimental group.  The symbol $\Sigma$ denotes the total count across experimental groups for a given task.}
    \resizebox{\linewidth}{!}{
    \begin{tabular}{l|c c c c c|c c c c c} 
        \toprule
        Task & \multicolumn{5}{c|}{\# Files} & \multicolumn{5}{c}{\# Participants} \\
        (Code)  &  CTL & PD & PDM & AD & $\Sigma$ & CTL & PD & PDM & AD & $\Sigma$ \\ 
	\midrule
Copy Cube (CC) &	38 & 25 & 14 & 39 & 116 & 36 & 22 & 12 & 20 & 90 \\
Copy Image (CI) &	42 & 37 & 16 & 42 & 137 & 39 & 34 & 14 & 21 & 108 \\
Copy Image Memory (CM) &	41 & 33 & 15 & 21 & 110 & 38 & 30 & 13 & 13 & 94 \\
Copy Read Text (CR) &	42 & 28 & 16 & 42 & 128 & 39 & 25 & 13 & 21 & 98 \\
Copy Text (CT) &	42 & 34 & 15 & 44 & 135 & 39 & 31 & 13 & 20 & 103 \\
Draw Clock (DC) &	43 & 27 & 15 & 41 & 126 & 40 & 24 & 13 & 20 & 97 \\
Freewrite (F) &	43 & 35 & 16 & 46 & 140 & 41 & 32 & 14 & 21 & 108 \\
Numbers (N) &	45 & 35 & 17 & 34 & 131 & 41 & 33 & 15 & 19 & 108 \\
Point Dominant (PD) &	44 & 36 & 16 & 43 & 139 & 41 & 32 & 13 & 21 & 107 \\
Point Non-Dominant (PN) &	43 & 36 & 17 & 43 & 139 & 40 & 33 & 14 & 21 & 108 \\
Point Left (PL) &	44 & 36 & 17 & 43 & 140 & 41 & 33 & 14 & 21 & 109 \\
Point Right (PR) &	43 & 36 & 16 & 43 & 138 & 40 & 32 & 13 & 21 & 106 \\
Point Sustained (PS) &	40 & 34 & 15 & 40 & 129 & 38 & 31 & 13 & 21 & 103 \\
Spiral Dominant (SD) &	41 & 36 & 16 & 45 & 138 & 38 & 33 & 14 & 21 & 106 \\
Spiral Non-Dominant (SN) &	41 & 36 & 15 & 45 & 137 & 40 & 33 & 13 & 21 & 107 \\
Spiral Left (SL) &	43 & 36 & 16 & 45 & 140 & 40 & 33 & 14 & 21 & 108 \\
Spiral Right (SR) &	39 & 36 & 15 & 45 & 135 & 38 & 33 & 13 & 21 & 105 \\
Spiral Pataka (SP) &	43 & 32 & 16 & 44 & 135 & 40 & 29 & 14 & 20 & 103 \\
\midrule
Total &	588 & 464 & 219 & 569 & 1840 & 42 & 35 & 15 & 21 & 113 \\
    \bottomrule
    \end{tabular}}
    \label{tab:files}
\end{table*}

\begin{table*}[ht]
    \centering
    \caption{Correlation of all the task-agnostic metrics with the MoCa score, UPDRS-III score and all its sub-components. 'n.s.' Is used for non significant features, $*$ for p-values$<0.05$, $**$ for p-values$<0.01$, $***$ for p-values$<0.001$, and $****$ for p-values$<0.0001$. }
    \resizebox{\linewidth}{!}{
    \begin{tabular}{l|c|c|c c c c c c c c c c c c c c c}
        \toprule
Feature & MoCa~\cite{julayanont2017montreal}& UPDRS-III~\cite{movement2003unified}& Rigidity& Upper& Lower & Arising & Gait& Freeze & Postural & Posture& Gait and & Kinetic & Postural & Resting & Global & Non-tremor \\
name & & & & bradykinesia & bradykinesia & from chair & & of gait & stability & & posture composite & tremor & tremor & tremor & tremor & components \\
\midrule
in-air duration	&	**** & n.s. & n.s. & n.s. & n.s. & n.s. & * & n.s. & n.s. & n.s. & n.s. & n.s. & n.s. & n.s. & ** & n.s.\\
in-air avg. X speed	&	* & n.s. & n.s. & n.s. & n.s. & n.s. & n.s. & n.s. & n.s. & n.s. & n.s. & * & ** & n.s. & * & n.s.\\
in-air std X speed	&	* & n.s. & n.s. & n.s. & n.s. & n.s. & n.s. & n.s. & n.s. & n.s. & n.s. & n.s. & * & n.s. & * & n.s.\\
in-air amp X speed	&	n.s. & n.s. & n.s. & n.s. & n.s. & n.s. & n.s. & * & n.s. & n.s. & n.s. & n.s. & n.s. & n.s. & n.s. & n.s.\\
in-air avg. X acc.	&	* & n.s. & n.s. & n.s. & n.s. & n.s. & n.s. & n.s. & n.s. & n.s. & n.s. & n.s. & * & n.s. & * & n.s.\\
in-air std X acc.	&	* & n.s. & n.s. & n.s. & n.s. & n.s. & n.s. & n.s. & n.s. & n.s. & n.s. & n.s. & n.s. & n.s. & * & n.s.\\
in-air amp X acc.	&	* & n.s. & n.s. & n.s. & n.s. & n.s. & n.s. & * & n.s. & n.s. & n.s. & n.s. & n.s. & n.s. & n.s. & n.s.\\
in-air avg. Y speed	&	* & n.s. & n.s. & n.s. & n.s. & ** & * & n.s. & n.s. & n.s. & n.s. & n.s. & n.s. & n.s. & * & n.s.\\
in-air std Y speed	&	n.s. & n.s. & n.s. & n.s. & n.s. & ** & * & n.s. & n.s. & n.s. & n.s. & n.s. & n.s. & n.s. & * & *\\
in-air amp Y speed	&	* & n.s. & n.s. & n.s. & n.s. & * & n.s. & n.s. & n.s. & n.s. & n.s. & n.s. & n.s. & n.s. & * & n.s.\\
in-air avg. Y acc.	&	* & n.s. & n.s. & n.s. & n.s. & *** & n.s. & n.s. & n.s. & n.s. & n.s. & n.s. & n.s. & n.s. & * & n.s.\\
in-air std Y acc.	&	* & n.s. & n.s. & n.s. & n.s. & * & n.s. & n.s. & n.s. & n.s. & n.s. & n.s. & n.s. & n.s. & * & n.s.\\
in-air amp Y acc.	&	** & n.s. & n.s. & n.s. & n.s. & * & n.s. & n.s. & n.s. & n.s. & n.s. & n.s. & n.s. & n.s. & * & n.s.\\
in-air avg. speed	&	**** & n.s. & n.s. & n.s. & n.s. & ** & n.s. & n.s. & n.s. & n.s. & n.s. & * & n.s. & n.s. & * & n.s.\\
in-air std speed	&	**** & n.s. & n.s. & n.s. & n.s. & n.s. & n.s. & n.s. & n.s. & n.s. & n.s. & * & n.s. & n.s. & * & n.s.\\
in-air amp speed	&	**** & n.s. & n.s. & n.s. & n.s. & n.s. & n.s. & n.s. & n.s. & n.s. & n.s. & * & n.s. & n.s. & * & n.s.\\
in-air avg. acc.	&	**** & n.s. & n.s. & n.s. & n.s. & * & n.s. & n.s. & n.s. & n.s. & n.s. & * & n.s. & n.s. & * & n.s.\\
in-air std acc.	&	**** & n.s. & n.s. & n.s. & n.s. & n.s. & n.s. & n.s. & n.s. & n.s. & n.s. & * & n.s. & n.s. & * & n.s.\\
in-air amp acc.	&	**** & n.s. & n.s. & n.s. & n.s. & n.s. & n.s. & n.s. & n.s. & n.s. & n.s. & * & n.s. & n.s. & n.s. & n.s.\\
in-air std angular speed	&	** & n.s. & n.s. & n.s. & n.s. & n.s. & n.s. & n.s. & n.s. & n.s. & n.s. & n.s. & n.s. & n.s. & n.s. & n.s.\\
in-air avg. angular speed	&	** & n.s. & n.s. & n.s. & n.s. & n.s. & * & n.s. & n.s. & n.s. & n.s. & * & * & n.s. & * & n.s.\\
in-air std angular acc.	&	** & n.s. & n.s. & n.s. & n.s. & n.s. & n.s. & n.s. & n.s. & n.s. & n.s. & n.s. & n.s. & n.s. & n.s. & n.s.\\
in-air avg. angular acc.	&	** & n.s. & n.s. & n.s. & n.s. & n.s. & * & n.s. & n.s. & n.s. & n.s. & * & * & n.s. & n.s. & n.s.\\
in-air shannon entropy	&	**** & n.s. & n.s. & n.s. & n.s. & * & n.s. & n.s. & n.s. & n.s. & n.s. & n.s. & n.s. & n.s. & * & n.s.\\
\midrule
on-tablet duration	&	*** & n.s. & n.s. & ** & n.s. & n.s. & * & n.s. & n.s. & n.s. & n.s. & n.s. & n.s. & n.s. & ** & n.s.\\
on-tablet avg. X speed	&	** & n.s. & n.s. & n.s. & * & * & * & n.s. & n.s. & n.s. & n.s. & n.s. & n.s. & n.s. & * & n.s.\\
on-tablet std X speed	&	** & n.s. & n.s. & n.s. & * & * & n.s. & n.s. & * & n.s. & n.s. & * & n.s. & n.s. & n.s. & n.s.\\
on-tablet amp X speed	&	n.s. & n.s. & n.s. & n.s. & ** & n.s. & n.s. & ** & ** & n.s. & n.s. & * & n.s. & n.s. & n.s. & n.s.\\
on-tablet avg. X acc.	&	n.s. & n.s. & n.s. & n.s. & * & * & * & n.s. & n.s. & n.s. & n.s. & * & n.s. & n.s. & * & n.s.\\
on-tablet std X acc.	&	n.s. & n.s. & n.s. & n.s. & * & n.s. & n.s. & ** & * & n.s. & n.s. & * & n.s. & n.s. & * & n.s.\\
on-tablet amp X acc.	&	n.s. & n.s. & n.s. & n.s. & * & n.s. & n.s. & ** & n.s. & n.s. & n.s. & * & n.s. & n.s. & n.s. & n.s.\\
on-tablet avg. Y speed	&	n.s. & n.s. & n.s. & n.s. & * & * & * & n.s. & n.s. & n.s. & n.s. & n.s. & n.s. & n.s. & n.s. & n.s.\\
on-tablet std Y speed	&	* & n.s. & n.s. & n.s. & * & * & * & n.s. & * & n.s. & n.s. & n.s. & n.s. & n.s. & n.s. & n.s.\\
on-tablet amp Y speed	&	n.s. & n.s. & n.s. & n.s. & * & n.s. & * & * & * & n.s. & n.s. & n.s. & n.s. & n.s. & n.s. & n.s.\\
on-tablet avg. Y acc.	&	n.s. & n.s. & n.s. & n.s. & * & n.s. & * & n.s. & n.s. & n.s. & n.s. & * & n.s. & n.s. & * & n.s.\\
on-tablet std Y acc.	&	n.s. & n.s. & n.s. & n.s. & ** & n.s. & n.s. & ** & n.s. & n.s. & n.s. & n.s. & n.s. & n.s. & n.s. & n.s.\\
on-tablet amp Y acc.	&	n.s. & n.s. & n.s. & n.s. & ** & n.s. & n.s. & ** & n.s. & n.s. & n.s. & n.s. & n.s. & n.s. & n.s. & n.s.\\
on-tablet avg. speed	&	n.s. & n.s. & n.s. & n.s. & *** & * & * & n.s. & ** & n.s. & n.s. & * & n.s. & n.s. & * & n.s.\\
on-tablet std speed	&	n.s. & n.s. & n.s. & n.s. & ** & n.s. & * & * & n.s. & n.s. & n.s. & n.s. & n.s. & * & * & n.s.\\
on-tablet amp speed	&	* & n.s. & n.s. & n.s. & ** & n.s. & n.s. & * & n.s. & n.s. & n.s. & n.s. & n.s. & n.s. & * & n.s.\\
on-tablet avg. acc.	&	n.s. & n.s. & n.s. & n.s. & *** & n.s. & * & * & ** & n.s. & n.s. & * & n.s. & n.s. & * & n.s.\\
on-tablet std acc.	&	n.s. & n.s. & n.s. & n.s. & ** & n.s. & n.s. & * & n.s. & n.s. & n.s. & * & n.s. & * & * & n.s.\\
on-tablet amp acc.	&	n.s. & n.s. & n.s. & n.s. & ** & n.s. & n.s. & ** & n.s. & n.s. & n.s. & n.s. & n.s. & n.s. & n.s. & n.s.\\
on-tablet std angular speed	&	* & n.s. & n.s. & n.s. & ** & n.s. & n.s. & n.s. & n.s. & n.s. & n.s. & n.s. & n.s. & n.s. & n.s. & n.s.\\
on-tablet avg. angular speed	&	**** & n.s. & n.s. & n.s. & n.s. & n.s. & n.s. & n.s. & n.s. & n.s. & n.s. & n.s. & n.s. & n.s. & n.s. & n.s.\\
on-tablet std angular acc.	&	n.s. & n.s. & n.s. & n.s. & ** & n.s. & n.s. & * & n.s. & n.s. & n.s. & n.s. & n.s. & n.s. & n.s. & n.s.\\
on-tablet avg. angular acc.	&	n.s. & n.s. & n.s. & n.s. & ** & n.s. & n.s. & n.s. & n.s. & n.s. & n.s. & n.s. & ** & n.s. & * & n.s.\\
on-tablet avg. dP/dt	&	n.s. & n.s. & n.s. & n.s. & * & n.s. & * & n.s. & n.s. & n.s. & n.s. & * & n.s. & * & ** & n.s.\\
on-tablet std dP/dt	&	n.s. & n.s. & n.s. & n.s. & ** & n.s. & * & n.s. & n.s. & n.s. & n.s. & * & n.s. & * & * & n.s.\\
on-tablet amp dP/dt	&	n.s. & n.s. & n.s. & n.s. & ** & n.s. & n.s. & * & n.s. & n.s. & n.s. & n.s. & n.s. & n.s. & n.s. & n.s.\\
on-tablet avg. d2P/dt2	&	n.s. & n.s. & n.s. & n.s. & ** & n.s. & * & n.s. & n.s. & n.s. & n.s. & * & n.s. & * & * & n.s.\\
on-tablet std d2P/dt2	&	n.s. & n.s. & n.s. & n.s. & ** & n.s. & * & n.s. & n.s. & n.s. & n.s. & * & n.s. & * & * & n.s.\\
on-tablet amp d2P/dt2	&	n.s. & n.s. & n.s. & n.s. & ** & n.s. & n.s. & * & n.s. & n.s. & n.s. & n.s. & n.s. & n.s. & n.s. & n.s.\\
on-tablet shannon entropy	&	*** & n.s. & n.s. & ** & * & n.s. & n.s. & n.s. & n.s. & n.s. & n.s. & n.s. & n.s. & n.s. & n.s. & n.s.\\
    \bottomrule
    \end{tabular}}
    \label{tab:significance_full}
\end{table*}

\begin{table*}[ht]
\centering
\caption{Table of classification performances for the Random Forest and Multi Layer Perceptron classifiers.}
\label{tab:classification}
\resizebox{\linewidth}{!}{
\begin{tabular}{|ccccccccccccccccccc|}
\hline
\multicolumn{1}{|c|}{\multirow{2}{*}{\textbf{Task}}} & \multicolumn{3}{c|}{\textbf{AD vs PD}} & \multicolumn{3}{c|}{\textbf{PD vs PDM}} & \multicolumn{3}{c|}{\textbf{AD vs CTL}} & \multicolumn{3}{c|}{\textbf{AD vs PDM}} & \multicolumn{3}{c|}{\textbf{PD vs CTL}} & \multicolumn{3}{c|}{\textbf{PDM vs CTL}} \\ \cline{2-19} 
\multicolumn{1}{|c|}{} & \multicolumn{1}{c|}{ACC} & \multicolumn{1}{c|}{F1} & \multicolumn{1}{c|}{AUC} & \multicolumn{1}{c|}{ACC} & \multicolumn{1}{c|}{F1} & \multicolumn{1}{c|}{AUC} & \multicolumn{1}{c|}{ACC} & \multicolumn{1}{c|}{F1} & \multicolumn{1}{c|}{AUC} & \multicolumn{1}{c|}{ACC} & \multicolumn{1}{c|}{F1} & \multicolumn{1}{c|}{AUC} & \multicolumn{1}{c|}{ACC} & \multicolumn{1}{c|}{F1} & \multicolumn{1}{c|}{AUC} & \multicolumn{1}{c|}{ACC} & \multicolumn{1}{c|}{F1} & AUC \\ \hline
\multicolumn{19}{|c|}{Random Forest} \\ \hline
\multicolumn{1}{|c|}{Copy Cube} & 64 & 63 & \multicolumn{1}{c|}{73} & 53 & 52 & \multicolumn{1}{c|}{58} & 63 & 64 & \multicolumn{1}{c|}{70} & 67 & 65 & \multicolumn{1}{c|}{69} & 64 & 63 & \multicolumn{1}{c|}{66} & 65 & 62 & 55 \\ \cline{1-1}
\multicolumn{1}{|c|}{Copy Image} & 83 & 83 & \multicolumn{1}{c|}{89} & \textit{53} & 46 & \multicolumn{1}{c|}{62} & 79 & 79 & \multicolumn{1}{c|}{84} & 78 & 78 & \multicolumn{1}{c|}{80} & 52 & 52 & \multicolumn{1}{c|}{49} & 58 & 55 & 64 \\ \cline{1-1}
\multicolumn{1}{|c|}{Copy Image Memory} & 77 & 77 & \multicolumn{1}{c|}{82} & 57 & 55 & \multicolumn{1}{c|}{56} & 62 & 62 & \multicolumn{1}{c|}{60} & 64 & 64 & \multicolumn{1}{c|}{65} & \textbf{69} & \textbf{69} & \multicolumn{1}{c|}{\textbf{75}} & 62 & 59 & 50 \\ \cline{1-1}
\multicolumn{1}{|c|}{Copy Read Text} & 72 & 71 & \multicolumn{1}{c|}{78} & 57 & 56 & \multicolumn{1}{c|}{54} & 63 & 63 & \multicolumn{1}{c|}{70} & 79 & 78 & \multicolumn{1}{c|}{\textbf{82}} & 65 & 63 & \multicolumn{1}{c|}{66} & 67 & 64 & 79 \\ \cline{1-1}
\multicolumn{1}{|c|}{Copy Text} & 82 & 82 & \multicolumn{1}{c|}{90} & \textbf{77} & \textbf{76} & \multicolumn{1}{c|}{\textbf{77}} & 79 & 79 & \multicolumn{1}{c|}{\textit{85}} & 77 & 75 & \multicolumn{1}{c|}{\textit{81}} & 61 & 61 & \multicolumn{1}{c|}{61} & 64 & 60 & 64 \\ \cline{1-1}
\multicolumn{1}{|c|}{Draw Clock} & 78 & 77 & \multicolumn{1}{c|}{88} & 50 & 49 & \multicolumn{1}{c|}{52} & 77 & 78 & \multicolumn{1}{c|}{\textit{88}} & 79 & 78 & \multicolumn{1}{c|}{68} & 60 & 60 & \multicolumn{1}{c|}{56} & 64 & 64 & 67 \\ \cline{1-1}
\multicolumn{1}{|c|}{Freewrite} & 73 & 73 & \multicolumn{1}{c|}{73} & 59 & 56 & \multicolumn{1}{c|}{\textit{70}} & 71 & 71 & \multicolumn{1}{c|}{77} & 79 & 79 & \multicolumn{1}{c|}{78} & 59 & 58 & \multicolumn{1}{c|}{55} & \textit{75} & \textit{74} & \textbf{79} \\ \cline{1-1}
\multicolumn{1}{|c|}{Numbers} & 66 & 65 & \multicolumn{1}{c|}{70} & 57 & 49 & \multicolumn{1}{c|}{46} & 68 & 68 & \multicolumn{1}{c|}{73} & 51 & 47 & \multicolumn{1}{c|}{48} & 50 & 50 & \multicolumn{1}{c|}{52} & 57 & 50 & 55 \\ \cline{1-1}
\multicolumn{1}{|c|}{Point Dominant} & \textit{88} & \textit{88} & \multicolumn{1}{c|}{\textit{93}} & 43 & 41 & \multicolumn{1}{c|}{39} & \textit{83} & \textit{82} & \multicolumn{1}{c|}{84} & \textit{82} & \textit{82} & \multicolumn{1}{c|}{80} & 62 & 62 & \multicolumn{1}{c|}{\textit{62}} & 59 & 57 & 53 \\ \cline{1-1}
\multicolumn{1}{|c|}{Point Non-Dominant} & \textit{88} & \textit{88} & \multicolumn{1}{c|}{\textbf{95}} & \textit{65} & 64 & \multicolumn{1}{c|}{56} & \textbf{84} & \textbf{83} & \multicolumn{1}{c|}{\textit{89}} & 78 & 76 & \multicolumn{1}{c|}{73} & \textit{68} & \textit{68} & \multicolumn{1}{c|}{\textit{71}} & 67 & 66 & 67 \\ \cline{1-1}
\multicolumn{1}{|c|}{Point Sustained} & \textbf{89} & \textbf{89} & \multicolumn{1}{c|}{\textit{93}} & 63 & 59 & \multicolumn{1}{c|}{51} & \textit{81} & \textit{81} & \multicolumn{1}{c|}{\textbf{90}} & \textbf{83} & \textbf{82} & \multicolumn{1}{c|}{79} & 65 & 65 & \multicolumn{1}{c|}{66} & 62 & 62 & 57 \\ \cline{1-1}
\multicolumn{1}{|c|}{Spiral Dominant} & 72 & 71 & \multicolumn{1}{c|}{71} & \textit{60} & 55 & \multicolumn{1}{c|}{56} & 68 & 68 & \multicolumn{1}{c|}{68} & 73 & 70 & \multicolumn{1}{c|}{58} & 49 & 48 & \multicolumn{1}{c|}{56} & 69 & 67 & 64 \\ \cline{1-1}
\multicolumn{1}{|c|}{Spiral Non-Dominant} & 63 & 62 & \multicolumn{1}{c|}{63} & \textit{68} & \textit{67} & \multicolumn{1}{c|}{68} & 70 & 69 & \multicolumn{1}{c|}{63} & 75 & 74 & \multicolumn{1}{c|}{67} & 45 & 43 & \multicolumn{1}{c|}{41} & 64 & 62 & 59 \\ \cline{1-1}
\multicolumn{1}{|c|}{Spiral PaTaKa} & 68 & 66 & \multicolumn{1}{c|}{69} & 66 & 65 & \multicolumn{1}{c|}{46} & 58 & 57 & \multicolumn{1}{c|}{52} & 73 & 71 & \multicolumn{1}{c|}{71} & 60 & 59 & \multicolumn{1}{c|}{62} & \textbf{79} & \textbf{78} & \textit{69} \\ \hline
\multicolumn{19}{|c|}{Multi Layer Perceptron} \\ \hline
\multicolumn{1}{|c|}{Copy Cube} & \textit{78} & \textit{78} & \multicolumn{1}{c|}{81} & 53 & 51 & \multicolumn{1}{c|}{52} & 56 & 56 & \multicolumn{1}{c|}{57} & 69 & 69 & \multicolumn{1}{c|}{69} & 57 & 57 & \multicolumn{1}{c|}{59} & 52 & 53 & 51 \\ \cline{1-1}
\multicolumn{1}{|c|}{Copy Image} & \textit{84} & \textit{84} & \multicolumn{1}{c|}{\textit{90}} & 67 & 63 & \multicolumn{1}{c|}{56} & 72 & 72 & \multicolumn{1}{c|}{82} & 69 & 68 & \multicolumn{1}{c|}{67} & 59 & 59 & \multicolumn{1}{c|}{62} & 49 & 48 & 48 \\ \cline{1-1}
\multicolumn{1}{|c|}{Copy Image Memory} & 65 & 64 & \multicolumn{1}{c|}{76} & 57 & 56 & \multicolumn{1}{c|}{59} & 55 & 55 & \multicolumn{1}{c|}{51} & 56 & 56 & \multicolumn{1}{c|}{66} & 57 & 57 & \multicolumn{1}{c|}{64} & 56 & 57 & 51 \\ \cline{1-1}
\multicolumn{1}{|c|}{Copy Read Text} & 65 & 63 & \multicolumn{1}{c|}{66} & 32 & 33 & \multicolumn{1}{c|}{33} & 66 & 66 & \multicolumn{1}{c|}{73} & 78 & 77 & \multicolumn{1}{c|}{87} & 61 & 59 & \multicolumn{1}{c|}{60} & 69 & 66 & \textbf{79} \\ \cline{1-1}
\multicolumn{1}{|c|}{Copy Text} & 77 & \textit{78} & \multicolumn{1}{c|}{\textit{84}} & 65 & 63 & \multicolumn{1}{c|}{\textit{69}} & 73 & 73 & \multicolumn{1}{c|}{78} & \textbf{86} & \textit{85} & \multicolumn{1}{c|}{\textbf{92}} & 58 & 58 & \multicolumn{1}{c|}{62} & 71 & 71 & 66 \\ \cline{1-1}
\multicolumn{1}{|c|}{Draw Clock} & 73 & 72 & \multicolumn{1}{c|}{79} & 59 & 59 & \multicolumn{1}{c|}{51} & 79 & 79 & \multicolumn{1}{c|}{\textbf{88}} & 77 & 76 & \multicolumn{1}{c|}{69} & 54 & 54 & \multicolumn{1}{c|}{61} & 52 & 53 & 50 \\ \cline{1-1}
\multicolumn{1}{|c|}{Freewrite} & 60 & 59 & \multicolumn{1}{c|}{63} & 59 & 56 & \multicolumn{1}{c|}{50} & 62 & 62 & \multicolumn{1}{c|}{72} & 75 & 75 & \multicolumn{1}{c|}{77} & 45 & 45 & \multicolumn{1}{c|}{44} & 73 & 72 & \textit{75} \\ \cline{1-1}
\multicolumn{1}{|c|}{Numbers} & 46 & 46 & \multicolumn{1}{c|}{44} & 39 & 38 & \multicolumn{1}{c|}{40} & 56 & 56 & \multicolumn{1}{c|}{70} & 45 & 45 & \multicolumn{1}{c|}{62} & 41 & 41 & \multicolumn{1}{c|}{46} & 57 & 55 & 53 \\ \cline{1-1}
\multicolumn{1}{|c|}{Point Dominant} & \textbf{88} & \textbf{88} & \multicolumn{1}{c|}{\textbf{92}} & 57 & 53 & \multicolumn{1}{c|}{57} & 81 & 81 & \multicolumn{1}{c|}{84} & 74 & 71 & \multicolumn{1}{c|}{75} & \textit{62} & \textit{62} & \multicolumn{1}{c|}{\textbf{73}} & 64 & 62 & 65 \\ \cline{1-1}
\multicolumn{1}{|c|}{Point Non-Dominant} & 83 & 83 & \multicolumn{1}{c|}{89} & 57 & 55 & \multicolumn{1}{c|}{57} & \textit{84} & \textit{83} & \multicolumn{1}{c|}{82} & 75 & 72 & \multicolumn{1}{c|}{72} & 59 & 58 & \multicolumn{1}{c|}{\textit{68}} & 64 & 64 & 65 \\ \cline{1-1}
\multicolumn{1}{|c|}{Point Sustained} & \textbf{88} & \textbf{88} & \multicolumn{1}{c|}{91} & 65 & 63 & \multicolumn{1}{c|}{\textbf{67}} & \textbf{87} & \textbf{87} & \multicolumn{1}{c|}{\textit{85}} & \textit{81} & \textbf{81} & \multicolumn{1}{c|}{\textit{80}} & \textbf{65} & \textbf{65} & \multicolumn{1}{c|}{\textit{68}} & 65 & 64 & 59 \\ \cline{1-1}
\multicolumn{1}{|c|}{Spiral Dominant} & 67 & 66 & \multicolumn{1}{c|}{74} & 47 & 42 & \multicolumn{1}{c|}{38} & 59 & 58 & \multicolumn{1}{c|}{58} & 78 & 75 & \multicolumn{1}{c|}{51} & 56 & 56 & \multicolumn{1}{c|}{55} & \textit{74} & \textit{74} & 64 \\ \cline{1-1}
\multicolumn{1}{|c|}{Spiral Non-Dominant} & 61 & 59 & \multicolumn{1}{c|}{60} & \textbf{75} & \textbf{73} & \multicolumn{1}{c|}{59} & 69 & 68 & \multicolumn{1}{c|}{69} & 73 & 66 & \multicolumn{1}{c|}{70} & 48 & 48 & \multicolumn{1}{c|}{49} & 57 & 51 & 52 \\ \cline{1-1}
\multicolumn{1}{|c|}{Spiral PaTaKa} & 66 & 66 & \multicolumn{1}{c|}{67} & \textit{70} & \textit{70} & \multicolumn{1}{c|}{54} & 50 & 50 & \multicolumn{1}{c|}{52} & 79 & 77 & \multicolumn{1}{c|}{71} & 58 & 58 & \multicolumn{1}{c|}{52} & \textbf{77} & \textbf{76} & 71 \\ \hline
\end{tabular}}
\end{table*}

\end{document}